\documentclass[12pt,reqno,a4paper]{article}
\usepackage[utf8]{inputenc}
\usepackage[british]{babel}
\usepackage{graphicx} 
\usepackage{amsmath}   
\usepackage{amssymb}   
\usepackage{amsfonts}
\usepackage{systeme,mathtools}
\usepackage{booktabs}
\usepackage{enumerate}
\usepackage{subcaption}
\usepackage{url}
\usepackage{mathdots}
\usepackage{multicol}
\usepackage{fancyhdr}
\usepackage{xcolor}
\usepackage{tikz-cd}

\newtheorem{theorem}{Theorem}
\newtheorem{proposition}{Proposition}

\newtheorem{definition}{Definition}

\newcommand{\bbZ}{{\mathbb Z}}

\newcommand{\bbC}{{\mathbb C}}
\newcommand{\bbP}{{\mathbb P}}

\def\t{\widetilde}

\title{Discrete Painlev\'e equations and pencils of quadrics in $\mathbb P^3$}
\author{Jaume Alonso, Yuri B. Suris, Kangning Wei}
\date{\small Institut für Mathematik, MA 7-1\\ Technische Universität Berlin, Str.\ des 17.\ Juni, 10623 Berlin, Germany\\
E-mail: alonso@math.tu-berlin.de, suris@math.tu-berlin.de, wei@math.tu-berlin.de}

\begin{document}

\maketitle

{\bf Abstract.} Discrete Painlev\'e equations constitute a famous class of integrable non-autonomous second order difference equations. A classification scheme proposed by Sakai interprets a discrete Painlev\'e equation as a birational map between generalized Halphen surfaces (surfaces obtained from $\mathbb P^1\times\mathbb P^1$ by blowing up at eight points). We propose a novel geometric interpretation of discrete Painlev\'e equations, where the family of generalized Halphen surfaces is replaced by a pencil of quadrics in $\mathbb P^3$. A discrete Painlev\'e equation is viewed as an autonomous  birational transformation of $\mathbb P^3$ that preserves the pencil and maps each quadric of the pencil to a different one, according to a M\"obius transformation of the pencil parameter. Thus, our scheme is based on the classification of pencils of quadrics in $\mathbb P^3$.

\section{Introduction}
\label{sect gen}

Discrete Painlev\'e equations belong to the central objects of interest in the theory of discrete integrable systems. Recall that continuous time Painlev\'e equations are second order nonlinear non-autonomous differential equations with the \emph{Painlev\'e property}, which is the absence of  moving singularities of solutions other than poles. Grammaticos, Ramani et.\ al.\ proposed a discrete version of the latter property called ``singularity confinement'', and found the first examples of second order nonlinear non-autonomous difference equations with this property, denoted by them as \emph{discrete Painlev\'e equations} \cite{GRP, RGH}. There followed a burst of activity on the subject summarized in \cite{GRreview}. A general classification scheme of discrete Painlev\'e equations was proposed by Sakai \cite{Sakai} and it is given a detailed exposition in the review paper by Kajiwara, Noumi and Yamada \cite{KNY}. A monographic account of discrete Painlev\'e equations is given by Joshi \cite{J}.

In the framework of Sakai's scheme, discrete Painlev\'e equations are birational maps between \emph{generalized Halphen surfaces} $X$. The latter can be realized as $\bbP^1\times\bbP^1$ blown up at eight points with the property that the anti-canonical divisor class $-K_X=2H_1+2H_2-\sum_{i=1}^8 E_i$ contains an effective divisor $D$ admitting a decomposition $D=\sum_{j=1}^m D_j$ of a \emph{canonical type}, where $D_j$ are irreducible effective divisors with
$$
 [D_j]\cdot [D_j]=-2, \quad  [D]\cdot [D_j]=0, \quad j=1,\ldots,m.
$$
Here $H_1$ and $H_2$ are the divisor classes of proper transforms of a generic vertical, resp. horizontal lines in $\bbP^1\times\bbP^1$, while $E_i$ is the total transform of the $i$-th blow-up. The scalar product in the Picard lattice ${\rm Pic}(X)=\bbZ H_1+\bbZ H_2+\sum_{i=1}^8 \bbZ E_i$ is given by the intersection number:
$$
H_1\cdot H_2 =1, \quad  E_i\cdot E_i=-1
$$
(all other scalar products among generators vanish). The matrix $( [D_i]\cdot [D_j])_{i,j=1}^m$ is the (negative of the) Cartan matrix of an affine root system $R(X)$, called the surface type of $X$. In particular, if $m=1$, so that the anti-canonical divisor class only contains irreducible effective divisors (irreducible curves of bidegree (2,2) in $\bbP^1\times \bbP^1$ passing through all eight blow-up points), one speaks about the surface type $A_0^{(1)}$.

Since the early days of the theory, discrete Painlev\'e equations were considered as non-autonomous versions (or modifications) of the so called \emph{QRT maps} \cite{QRT1, QRT2, QRT book}. The latter are birational maps of $\bbP^1\times \bbP^1$ defined as compositions of a vertical and a horizontal involutions generated by a \emph{pencil of biquadratic curves}. While in the old-style literature the non-autonomous modification was mainly introduced in an ad hoc way by allowing some coefficients of the map to become time-dependent, a more geometric version of this procedure was proposed in the framework of the Sakai's scheme by Carstea, Dzhamay and Takenawa \cite{CDT}. In their scheme, the de-automization of a given QRT map depends on the choice of one biquadratic curve of the pencil.

In the present paper, we propose an alternative view on discrete Painlev\'e equations, and simultaneously an alternative procedure for the de-autonomization of QRT maps. In our scheme, the surfaces on which discrete Painlev\'e equations act are quadrics of a pencil in $\bbP^3$. Our scheme can be described as follows.
\begin{enumerate}
\item Start with a pencil $\{C_\mu\}$ of biquadratic curves in $\bbP^1\times\bbP^1$ and the corresponding QRT map. Let $s_1,\ldots,s_8\in\bbP^1\times\bbP^1$ be the base points of this pencil. Lift $\{C_\mu\}$ to a pencil of quadrics $\{P_\mu\}$ in $\bbP^3$. The base curve of this pencil passes through the lifts $S_1,\ldots,S_8$ of the base points $s_1,\ldots,s_8$. 
\item Choose one distinguished biquadratic curve $C_\infty$ of the pencil, along with its lift to a quadric $P_\infty$.
\item Based on these data, construct the pencil of quadrics $\{Q_\lambda\}$  in $\bbP^3$ spanned by $Q_0=\{X_1X_2-X_3X_4=0\}$ and by $P_\infty$. Recall that $Q_0$ is nothing but the \emph{Segre embedding} of $\bbP^1\times\bbP^1$ to $\bbP^3$.  The base curve of the pencil $\{Q_\lambda\}$ is, by definition, the curve $Q_0\cap P_\infty$, which is the image of $C_\infty$ under the Segre embedding. The intersection of this curve with the base curve of the pencil $\{P_\mu\}$ consists exactly of the points $S_1,\ldots,S_8$.
\item Consider a \emph{3D QRT map} on the pencil $\{Q_\lambda\}$ defined by intersections of its generators with the quadrics $P_\mu$. Recall that the notion of 3D QRT maps was introduced in \cite{ASW} and that such a map preserves each quadric $Q_\lambda$ (and therefore the pencil parameter $\lambda$ serves as an integral of motion). On each quadric $Q_\lambda$, our map induces a QRT map which can be considered as a $\lambda$-deformation of the original QRT map.
\item Consider a birational map $L$ on $\bbP^3$ with the following properties.
\begin{itemize}
\item[a)] $L$ preserves the pencil $\{Q_\lambda\}$ and its base curve, and maps each $Q_\lambda$ to $Q_{\sigma(\lambda)}$, where $\sigma:\bbP^1\to\bbP^1$ is a M\"obius automorphism fixing the set 
\begin{equation}\label{Sing}
{\rm Sing}(Q):=\big\{\lambda\in\bbP^1: \;Q_\lambda\;{\rm is\;\; degenerate}\big\}.
\end{equation}
\item[b)] The maps $L\circ i_1,L\circ i_2$ have the same singularity confinement properties as the QRT involutions $i_1, i_2$.
\end{itemize} 
Then the map $(L\circ i_1)\circ(L\circ i_2)$ is declared to be the {\em 3D Painlev\'e map} corresponding to the de-autonomization of the QRT map along the fiber $C_\infty$.
\end{enumerate}

Let us point out the most distinctive feature of our picture. Our 3D Painlev\'e maps are {\em autonomous birational maps of $\mathbb P^3$}. Correspondingly, all fundamental ingredients: the distinguished fiber (the Segre lift of $C_\infty$), the base points $S_i$, and the blow-up points (which include some of $S_i$ and some of $L(S_i)$) {\em do not move} in the discrete time evolution. {\em Their movement is apparent} and is due to the change of the pencil adapted coordinates by transitioning from $Q_\lambda$ to $Q_{\sigma(\lambda)}$. This is the fundamental difference from constructions based on the birational action of the affine Weyl group on configurations of eight points in $\mathbb P^3$, cf. \cite{T}.
\smallskip

The structure of the paper is as follows. We start by recalling several general concepts necessary for our presentation, namely the notion of singularity confinement for birational maps (Section \ref{sect sing conf}), construction and basic properties of QRT maps (Section \ref{sect QRT}), a three-dimensional generalization of QRT maps introduced in \cite{ASW} (Section \ref{sect 3D QRT}), as well as a classical projective classification of pencil of quadrics in $\mathbb P^3$ (Section \ref{sect pencils}). Then, in Section \ref{sect 3D QRT for a fiber}, we describe in detail the points 3 and 4 of the scheme above, i.e., a construction of a 3D QRT map based on the choice of a biquadratic curve in the invariant fibration of a given QRT map. Finally, the general part culminates in the discussion of the notion of the Painlev\'e deformation in Section \ref{sect L}. There follow seven Sections \ref{sect case ix}--\ref{sect case vii} contaning a detailed elaboration of our scheme for seven (out of thirteen) projective classes of pencils of quadrics. These cases are characterized by the property that the characteristic polynomial of the pencil is a complete square, and, as a consequence, the generators of the pencil are rational functions on $\mathbb P^3$. 

In the present form, our scheme covers discrete Painlev\'e equations of the Sakai's scheme for all surface types below $A_0^{(1)}$. It does include a multiplicative and additive versions of $A_1^{(1)}$ (Sections \ref{sect case viii}, \ref{sect case vii}), however in a realization different from the standard one \cite{KNY}.  

Modifications of our scheme necessary to treat the remaining six cases (the multiplicative and the additive discrete Painlev\'e equations of the type $A_1^{(1)}$ in the standard realization, as well as the elliptic, the multiplicative and the additive equations of the type $A_0^{(1)}$) will be discussed in a subsequent paper \cite{ASWnets}. 

\subparagraph*{Acknowledgement.}
This research was supported by the DFG Collaborative Research Center TRR 109 ``Discretization in
Geometry and Dynamics''.


\section{Generalities: singularity confinement}
\label{sect sing conf}

For birational maps of $\mathbb P^N$, we will use the following basic notions and results \cite{H}, \cite{D}, \cite{DF}. Let such a map
$f: \mathbb{P}^N\dasharrow\mathbb{P}^N$ be given by $N+1$ homogeneous polynomials of one and the same degree $d$ without a non-trivial common factor,
\begin{equation}\label{eq: rational map}
f: \; [x_0:x_1:\ldots:x_N]\mapsto[X_0:X_1:\ldots:X_N].
\end{equation}
The number $d$ is called the {\em degree} of $f$, denoted by $\deg(f)$. The corresponding polynomial map $\tilde f$ of $\mathbb C^{N+1}$ is called a {\em minimal lift} of $f$. It is defined up to a constant factor.
To each birational map we associate:
\begin{itemize}
\item the {\em indeterminacy set} $\mathcal I(f)$ consisting of the points $ [x_0:x_1:\ldots:x_N]\in \mathbb{P}^N$ for which $X_0=X_1=\ldots=X_N=0$;  this is a variety of codimension at least 2;
\item the {\em critical set} $\mathcal C(f)$ consisting of the points $[x_0:x_1:\ldots:x_N]\in \mathbb{P}^N$ where $\det d\tilde f=0$; the latter equation of degree $(N+1)(d-1)$ defines a variety of codimension 1.
\end{itemize} 
There holds $\mathcal I(f)\subset \mathcal C(f)$. Away from $\mathcal C(f)$, the map $f$ acts biregularly. On the other hand, the image of $\mathcal C(f)\setminus \mathcal I(f)$ under the map $f$ belongs to $\mathcal I(f^{-1})$ (in particular, it is of codimension $\ge 2$). Loosely speaking, $f$ contracts (or blows down) hypersurfaces from $\mathcal C(f)$.

The further fate of the images of $\mathcal C(f)$ under iterates of $f$ is essential for the notion of singularity confinement.
This notion was originally introduced in \cite{GRP} as an integrability criterium, see its current status in \cite{MWRG}. An algebro-geometric interpretation of singularity confinement, which we adopt below, followed in \cite{BV}, where it was shown to be related to the phenomenon of the drop of degree of iterates $f^k$, which in turn is responsible for the drop of the dynamical degree of $f$. The drop of degree of $f^k$ happens if in all components of the polynomial map $\tilde f^k$ there appears a common polynomial factor. A geometric condition for this is the existence of degree lowering hypersurfaces.
\begin{definition}\label{def deg lowering}
A hypersurface $A=\{a=0\}\subset \mathcal C(f)$ (where $a(x_0,x_1,\ldots,x_N)$ is a homogeneous polynomial on $\mathbb C^{N+1}$) is called a  \emph{degree lowering hypersurface} for the map $f$ if $f^{k-1}(A)\subset \mathcal I(f)$ for some $k\in\mathbb N$.
\end{definition}
Indeed, in this case all components of $\tilde f^k$ vanish as soon as $a=0$, and therefore are divisible by $a$. As a corollary,  the dynamical degree of the map $f$, defined as
\begin{equation}\label{dyn deg}
\lambda(f):=\lim_{n\to\infty} (\deg(f^n))^{1/n},
\end{equation}
is strictly less than $\deg f$. The condition $\lambda(f)=1$, or, equivalently, the vanishing of the \emph{algebraic entropy} $h(f)=\log\lambda(f)$, is a popular definition of \emph{integrability} of a birational map $f$ (cf. \cite{BV}). In particular, one often refers to this definition (or criterium) when speaking about discrete Painlev\'e equations \cite{J}, \cite{KNY}.

For a degree lowering hypersurface $A\subset\mathcal C(f)$, with $f^k(A)\subset \mathcal I(f)$, we call the diagram
\begin{equation}\label{sing conf pattern}
A\stackrel{f}{\to} f(A)\stackrel{f}{\to} f^2(A)\stackrel{f}{\to} \ldots\stackrel{f}{\to} f^k(A)\stackrel{f}{\to} B
\end{equation}
a \emph{singularity confinement pattern}. In such a pattern, one should think of $f$ as blowing up $f^k(A)\subset \mathcal I(f)$ to a hypersurface $B\subset\mathcal C(f^{-1})$. In the present paper, we will not address the issue of regularizing the map $f$, i.e., lifting it to a blow-up variety $X$ so that the lift is algebraically stable (does not possess degree lowering hypersurfaces). According to a theorem by Diller and Favre \cite{DF}, this is always possible in dimension $N=2$. For a map with an algebraically stable lift, the dynamical degree $\lambda(f)$ can be computed as the spectral radius of the induced action of this lift on the Picard group ${\rm Pic}(X)$. All our examples here (in dimensions $N=2$ and $N=3$) possess algebraically stable lifts, moreover, \emph{all singularities are confined} in the sense that \emph{all components of the critical set are degree lowering, resulting in singularity confinement patterns} as in \eqref{sing conf pattern}.


\section{Generalities: QRT maps}
\label{sect QRT}

To quickly introduce QRT maps, consider a \emph{pencil of biquadratic curves}
$$
C_\mu=\Big\{(x,y)\in \bbC^2: C_\mu(x,y):=C_0(x,y)-\mu C_\infty(x,y)=0\Big\},
$$
where $C_0,C_\infty$ are two polynomials of bidegree (2,2). The \emph{base set} $\mathcal B$ of the pencil is defined as the set of points through which all curves of the pencil pass or, equivalently, as the intersection $\{C_0(x,y)=0\}\cap\{C_\infty(x,y)=0\}$. 
Through any point $(x_0,y_0)\not\in \mathcal B$, there passes exactly one curve of the pencil, defined by $\mu=\mu(x_0,y_0)=C_0(x_0,y_0)/C_\infty(x_0,y_0)$. Actually, we consider this pencil in a compactification $\bbP^1\times \bbP^1$ of $\bbC^2$. Then, $B$ consists of eight base points, counted with multiplicity, $\mathcal B=\{s_1,\ldots,s_8\}\subset\bbP^1\times\bbP^1$.

One defines the \emph{vertical switch} $i_1$ and the \emph{horizontal switch} $i_2$ as follows. For a given point $(x_0,y_0)\in\bbP^1\times \bbP^1 \setminus \mathcal B$, determine $\mu=\mu(x_0,y_0)$ as above. Then the vertical line $\{x=x_0\}$ intersects $C_\mu$ at exactly one further point $(x_0,y_1)$ which is defined to be $i_1(x_0,y_0)$; similarly, the horizontal line $\{y=y_0\}$ intersects $C_\mu$ at exactly one further point $(x_1,y_0)$ which is defined to be $i_2(x_0,y_0)$. The QRT map is defined as
$$
f=i_1\circ i_2.
$$
Each of the maps $i_1$, $i_2$ is a birational involution on $\bbP^1\times \bbP^1$ with indeterminacy set $\mathcal B$. Likewise, the QRT map $f$ is a (dynamically nontrivial) birational map on $\bbP^1\times \bbP^1$, having $\mu(x,y)=C_0(x,y)/C_\infty(x,y)$ as an integral of motion. A generic fiber $C_\mu$ is an elliptic curve, and $f$ acts on it as a shift with respect to the corresponding addition law. 

We briefly discuss singularity confinement patterns for QRT maps. 
\begin{itemize}

\item If a base point $s_i=(a_i,b_i)$ is the only base point on the line $\{x=a_i\}$ and the only base point on the line $\{y=b_i\}$, then it is an indeterminacy point for both involutions $i_1$, $i_2$. More precisely, $i_1$ blows down the line $\{x=a_i\}$ to the point $s_i$ (and, since it is an involution, blows up the point $s_i$ to the line  $\{x=a_i\}$). Likewise, $i_2$ blows down the line $\{y=b_i\}$ to the point $s_i$ and blows up the point $s_i$ to the line $\{y=b_i\}$. We say that the following \emph{short singularity confinement pattern} for the involutions $i_1$, $i_2$ happens:
\begin{equation}\label{sing conf QRT i1i2}
\{x=a_i\}\; \stackrel{i_1}{\rightarrow}\; s_i\; \stackrel{i_2}{\rightarrow}\; \{y=b_i\}.
\end{equation}
As a consequence, we have also a short singularity confinement pattern for the map $f=i_1\circ i_2$:
\begin{equation}\label{sing conf QRT f type 1}
i_2(\{x=a_i\}\; \stackrel{f}{\rightarrow}\; s_i\; \stackrel{f}{\rightarrow}\; i_1(\{y=b_i\}).
\end{equation}

\item If there are two base points $s_i=(a_i,b_i)$ and $s_j=(a_i,b_j)$ on the line $\{x=a_i\}$, then both are singularities for the involution $i_2$, being blown up to the corresponding lines $\{y=b_i\}$, resp. $\{y=b_j\}$. On the contrary, for the involution $i_1$ the line $\{x=a_i\}$ is invariant; $i_1$ induces a projective (M\"obius) involution on this line, which interchanges the points $s_i$ and $s_j$. We say that the following \emph{long singularity confinement patterns} for the involutions $i_1$, $i_2$ happen:
\begin{equation}\label{sing conf QRT i1i2 long y}
\{y=b_i\}\; \stackrel{i_2}{\rightarrow}\; s_i \; \stackrel{i_1}{\rightarrow}\; s_j \; \stackrel{i_2}{\rightarrow}\; \{y=b_j\}
\end{equation}
(and a similar one with the roles of $s_i$ and $s_j$ interchanged). As a consequence, we have also short singularity confinement patterns for the map $f=i_1\circ i_2$:
\begin{equation}\label{sing conf QRT f type 2}
\{y=b_i\}\; \stackrel{f}{\rightarrow}\; s_j\; \stackrel{f}{\rightarrow}\; i_1(\{y=b_j\})
\end{equation}
(and a similar one involving $s_i$). 

\item Analogously, if there are two base points $s_i=(a_i,b_i)$ and $s_j=(a_j,b_i)$ on the line $\{y=b_i\}$, then both are singularities for the involution $i_1$, being blown up to the corresponding lines $\{x=a_i\}$, resp. $\{x=a_j\}$. For the involution $i_2$ the line $\{y=b_i\}$ is invariant; it induces a projective (M\"obius) involution on this line, which interchanges the points $s_i$ and $s_j$. A \emph{long singularity confinement pattern} happens for the involutions $i_1$, $i_2$:
\begin{equation}\label{sing conf QRT i1i2 long x}
\{x=a_i\}\; \stackrel{i_1}{\rightarrow}\; s_i \; \stackrel{i_2}{\rightarrow}\; s_j \; \stackrel{i_1}{\rightarrow}\; \{x=a_j\}
\end{equation}
(and a similar one with the roles of $s_i$ and $s_j$ interchanged). As a consequence, a short singularity confinement pattern happens for the map $f=i_1\circ i_2$:
\begin{equation}\label{sing conf QRT f type 3}
i_2(\{x=a_i\})\; \stackrel{f}{\rightarrow}\; s_i\; \stackrel{f}{\rightarrow}\; \{x=a_j\}
\end{equation}
(and a similar one involving $s_j$). 
\end{itemize}
Summarizing, for the map $f$ there are eight short singularity confinement patterns, each of the base points participating in exactly one pattern. For the involutions $i_1$, $i_2$ one also has eight singularity confinement patterns, but some of them become long if the base points are in a special relative position. Of course, further degenerations are possible in case of further geometric specialties in the configuration of the base points, e.g., if there are infinitely near points among them.


\section{Generalities: three-dimensional QRT maps}
\label{sect 3D QRT}

On the way towards a 3D generalization of QRT maps, the first step is a translation of the construction just described to a Segre embedding of $\bbP^1\times\bbP^1$ as a quadric in $\bbP^3$:
\begin{equation}\label{Q0}
Q_0=\Big\{[X_1:X_2:X_3:X_4]: X_1X_2-X_3X_4=0\Big\}\subset \bbP^3.
\end{equation}
Thus, $Q_0$ is isomorphic to $\bbP^1\times\bbP^1$, via the \emph{Segre embedding}
\begin{equation}
\phi_0:\quad \bbP^1\times\bbP^1\ni \big([x_1:x_0],[y_1:y_0]\big)\,\mapsto \; [x_1y_0:x_0y_1:x_1y_1:x_0y_0]\in Q_0.
\end{equation}
Usually, we write this in the affine chart $\bbC\times\bbC$ of $\bbP^1\times\bbP^1$ as follows:
\begin{equation}
\phi_0:\quad \bbP^1\times\bbP^1\ni (x,y)\,\mapsto \;[x:y:xy:1]\in Q_0.
\end{equation}
The quadric $Q_0$, like any non-degenerate quadric in $\bbP^3$, admits two rulings such that any two lines of one ruling are skew and any line of one ruling intersects any line of the other ruling. Through each point $X\in Q_0$ there pass two straight lines, one of each of the two rulings, let us call them $\ell_1(X)$ and $\ell_2(X)$. More concretely, $\ell_1(X)$ can be described as $\{x={\rm const}\}$, while $\ell_2(X)$ can be described as $\{y={\rm const}\}$.

Now, any biquadratic curve in $\bbP^1\times \bbP^1$ with the equation
$$
C:\; \big\{a_{1}x^2y^2+a_{2}x^2y+a_{3}xy^2 
+a_{4}x^2+a_{5}y^2+a_6xy+a_7x+a_8y +a_9 =0\big\}
$$
can be identified with $Q_0\cap P$, where $P$ is the quadric in $\mathbb P^3$ with the equation
$$
P:\;\big\{ a_{1}X_3^2+a_{2}X_1X_3+a_{3}X_2X_3+a_{4}X_1^2 +a_{5}X_2^2+a_{6}X_3X_4+a_{7}X_1X_4 
   +a_{8}X_2X_4+a_{9}X_4^2 =0\big\}.
$$
We call the quadric $P$ the {\em Segre lift} of the biquadratic curve $C$.
Therefore, to a pencil of biquadratic curves $\{C_\mu\}$ in $\bbP^1\times \bbP^1$ there corresponds a pencil of quadrics $\{P_\mu\}$ in $\bbP^3$, their Segre lifts. The corresponding QRT map can be identified with $f=i_1\circ i_2$, where $i_1$, $i_2$ are involutions on $Q_0$ defined as follows. For a given point $X\in Q_0$, different from the lifts of the base points of the pencil $\{C_\mu\}$, let $\mu=\mu(X)$ be defined as the value of the pencil parameter for which $X\in Q_0\cap P_\mu$. Denote by $i_1(X)$, $i_2(X)$ the second intersection point of $\ell_1(X)$ with $P_\mu$, resp. the second intersection point of $\ell_2(X)$ with $P_\mu$.
\smallskip

Now we are in a position to give a three-dimensional generalization of the QRT construction. For this, consider a second pencil of quadrics $\{Q_\lambda\}$ in $\bbP^3$, and consider the QRT construction on each fiber $Q_\lambda$ individually.

\begin{definition} \label{def 3D QRT}
Given two pencils of quadrics $\{Q_\lambda\}$ and $\{P_\mu\}$, we define involutions $i_1, i_2:\bbP^3\to \bbP^3$ as follows: for a generic $X\in \bbP^3$ (not belonging to the base set of either pencil), determine $\lambda,\mu\in\bbP^1$ such that $X\in Q_\lambda\cap P_\mu$; then $i_1(X)$ is defined to be the second intersection point of the generator $\ell_1(X)$ of $Q_\lambda$ with $P_\mu$, and similarly $i_2(X)$ is the second intersection point of the generator $\ell_2(X)$ of $Q_\lambda$ with $P_\mu$. The 3D QRT map is defined as $f=i_1\circ i_2:\bbP^3\to \bbP^3$; it leaves all quadrics of both pencils invariant.
\end{definition}

The main problem with this definition is that the dependence of generators $\ell_1(X)$, $\ell_2(X)$ on the point $X$ can be non-rational.  This issue is the subject of the following section.

\section{ Generalities: pencils of quadrics}
\label{sect pencils}

Let $\{Q_\lambda\}_{\lambda\in \bbP^1}$ be a pencil of quadrics in $\bbP^3$, with $Q_\lambda=Q_0-\lambda Q_\infty$. Denote by $M_0,M_\infty\in {\rm Sym}_{4\times 4}(\bbC)$ symmetric matrices of the quadratic forms $Q_0,Q_\infty$, and set
$M_\lambda=M_0-\lambda M_\infty$. 

It is well known (see, e..g., \cite{CA}) that pencils of quadrics in $\bbP^3$ are classified, modulo complex congruence transformations, by the structure of the system of elementary divisors of $M_\lambda$, encoded in the so called \emph{Segre symbols}. Elementary divisors are powers of $\lambda-\lambda_k$ for $\lambda_k\in\,{\rm Sing}(Q):=\{\lambda\in\bbP^1: Q_\lambda\;\;{\rm is\;\; singular}\}$. The product of all elementary divisors is the characteristic polynomial
\begin{equation}
\Delta(\lambda)=\det(M_\lambda)=\det(M_0-\lambda M_\infty).
\end{equation}
The classification of pencils of quadrics in $\bbP^3$ modulo complex congruence transformations consists of the following thirteen classes:
\begin{itemize}

\item[(i)] \emph{Pencil of quadrics through a non-singular spatial quartic curve.}
\newline
Segre symbol $[1,1,1,1]$; $\Delta(\lambda)=(\lambda-\lambda_1)(\lambda-\lambda_2)(\lambda-\lambda_3)(\lambda-\lambda_4)$.

\item[(ii)] \emph{Pencil of quadrics through a nodal spatial quartic curve}.
\newline
Segre symbol $[2,1,1]$;  $\Delta(\lambda)=(\lambda-\lambda_1)^2(\lambda-\lambda_2)(\lambda-\lambda_3)$.

\item[(iii)] \emph{Pencil of quadrics through a cuspidal spatial quartic curve}. 
\newline 
Segre symbol $[3,1]$; $\Delta(\lambda)=(\lambda-\lambda_1)^3(\lambda-\lambda_2)$.

\item[(iv)] \emph{Pencil of quadrics through two non-coplanar conics sharing two points.} 
\newline 
Segre symbol $[(1,1),1,1]$; $\Delta(\lambda)=(\lambda-\lambda_1)^2(\lambda-\lambda_2)(\lambda-\lambda_3)$.

\item[(v)] \emph{Pencil of quadrics through two non-coplanar conics touching at a point.}
\newline 
Segre symbol $[(2,1),1]$; $\Delta(\lambda)=(\lambda-\lambda_1)^3(\lambda-\lambda_2)$.

\item[(vi)] \emph{Pencil of quadrics tangent along a non-degenerate conic.}
\newline 
Segre symbol  $[(1,1,1),1]$; $\Delta(\lambda)=(\lambda-\lambda_1)^3(\lambda-\lambda_2)$.

\item[(vii)] \emph{Pencil of quadrics through a twisted cubic and one of its chords.}
\newline 
Segre symbol $[2,2]$; $\Delta(\lambda)=(\lambda-\lambda_1)^2(\lambda-\lambda_2)^2$.

\item[(viii)] \emph{Pencil of quadrics through a twisted cubic and one of its tangents.}
\newline 
Segre symbol $[4]$; $\Delta(\lambda)=(\lambda-\lambda_1)^4$.

\item[(ix)] \emph{Pencil of quadrics through a conic and two coplanar lines through different points of the conic.}
\newline 
Segre symbol $[2,(1,1)]$; $\Delta(\lambda)=(\lambda-\lambda_1)^2(\lambda-\lambda_2)^2$.

\item[(x)] \emph{Pencil of quadrics through a conic and two lines meeting on the conic.}
\newline 
Segre symbol $[(3,1)]$; $\Delta(\lambda)=(\lambda-\lambda_1)^4$.

\item[(xi)] \emph{Pencil of quadrics through a skew quadrilateral.}
\newline 
Segre symbol $[(1,1),(1,1)]$; $\Delta(\lambda)=(\lambda-\lambda_1)^2(\lambda-\lambda_2)^2$.

\item[(xii)] \emph{Pencil of quadrics through three lines, tangent along one of them.}
\newline 
Segre symbol $[(2,2)]$; $\Delta(\lambda)=(\lambda-\lambda_1)^4$.

\item[(xiii)] \emph{Pencil of quadrics tangent along a pair of lines.}
\newline 
Segre symbol $[(2,1,1)]$; $\Delta(\lambda)=(\lambda-\lambda_1)^4$.
\end{itemize}
By a projective (M\"obius) transformation of $\lambda$, one can achieve $\lambda_1=\infty$, $\lambda_2=0$, $\lambda_3=1$. In the case (i), we have one module, the cross-ratio of $\lambda_1,\ldots,\lambda_4$. All other cases are exhausted by just one pencil, up to the projective transformations of $\mathbb P^3$ and M\"obius transformations of $\lambda$ (e.g., with the values of $\lambda_i$ just mentioned). 

\renewcommand\thesubfigure{\roman{subfigure}}
\begin{figure}[ht!]
     \centering
     \begin{subfigure}[b]{0.15\textwidth}
         \centering
         \includegraphics[width=\textwidth]{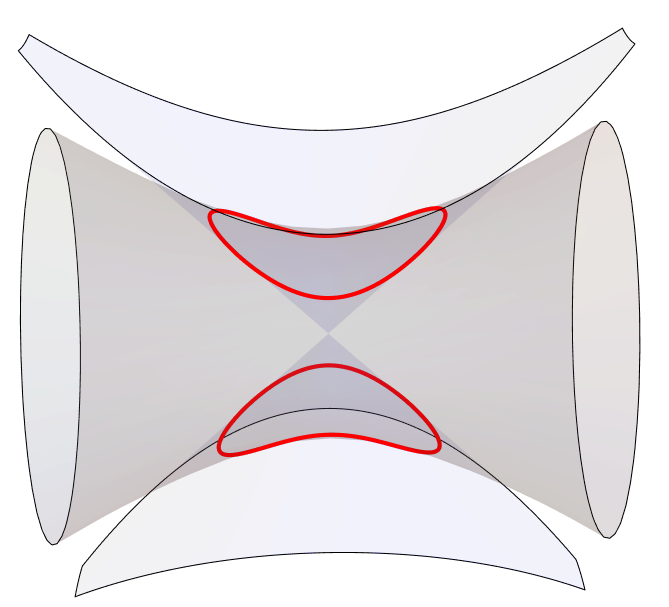}
         \caption{}
     \end{subfigure}
     \hspace{1cm}
     \begin{subfigure}[b]{0.21\textwidth}
         \centering
         \includegraphics[trim={0 0 20 30pt},width=0.9\textwidth]{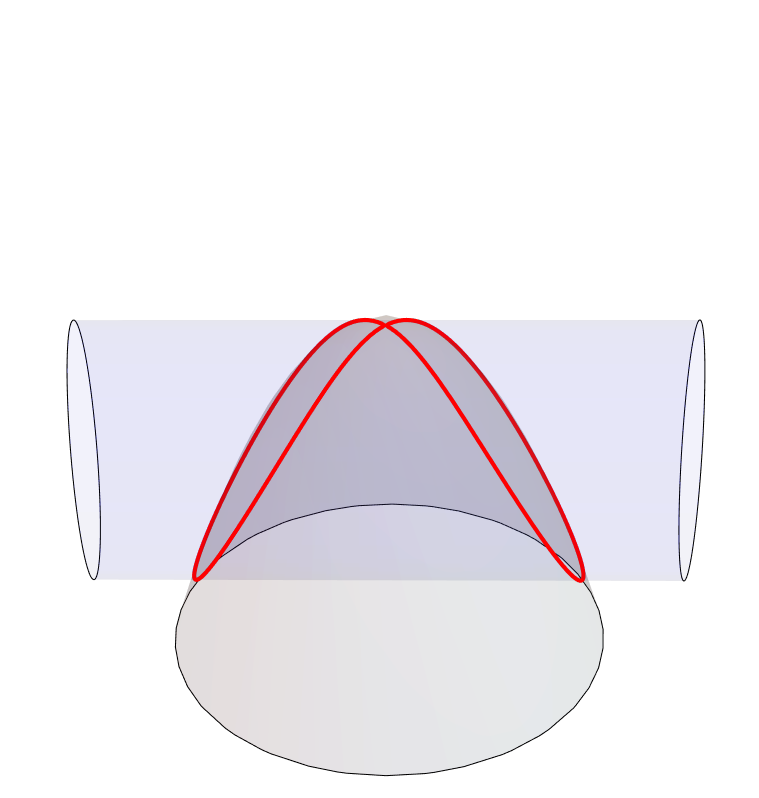}
         \caption{}
     \end{subfigure}
     \hspace{1cm}
     \begin{subfigure}[b]{0.225\textwidth}
         \centering
         \includegraphics[trim={0 30 0 60},width=0.9\textwidth]{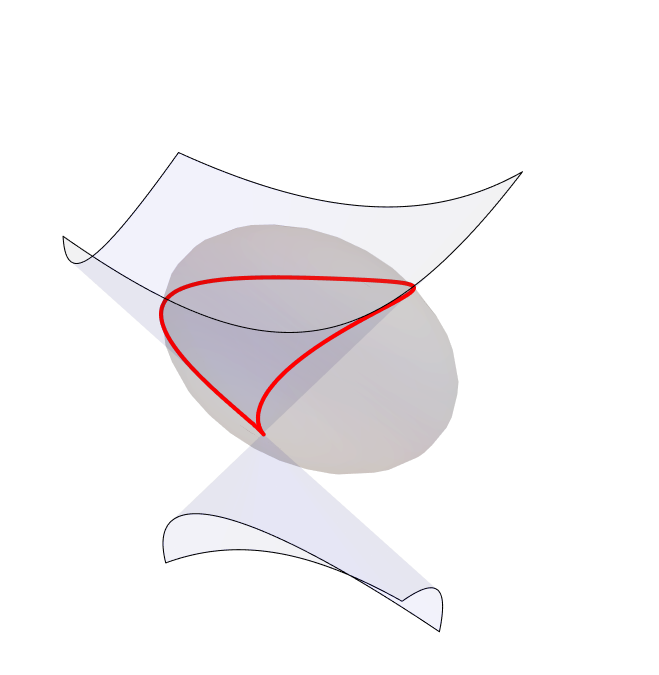}
         \caption{}
     \end{subfigure}\\
          \begin{subfigure}[b]{0.21\textwidth}
		\centering
         \includegraphics[trim={40 40 0 0},width=\textwidth]{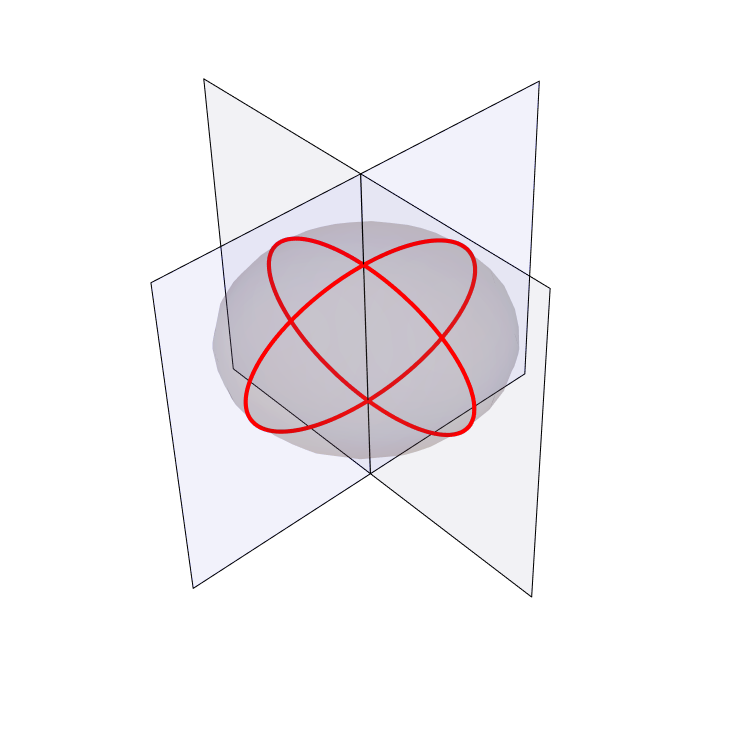}
         \caption{}
     \end{subfigure}
     \hspace{1cm}
     \begin{subfigure}[b]{0.225\textwidth}
         \includegraphics[trim={0 30 0 0},width=0.9\textwidth]{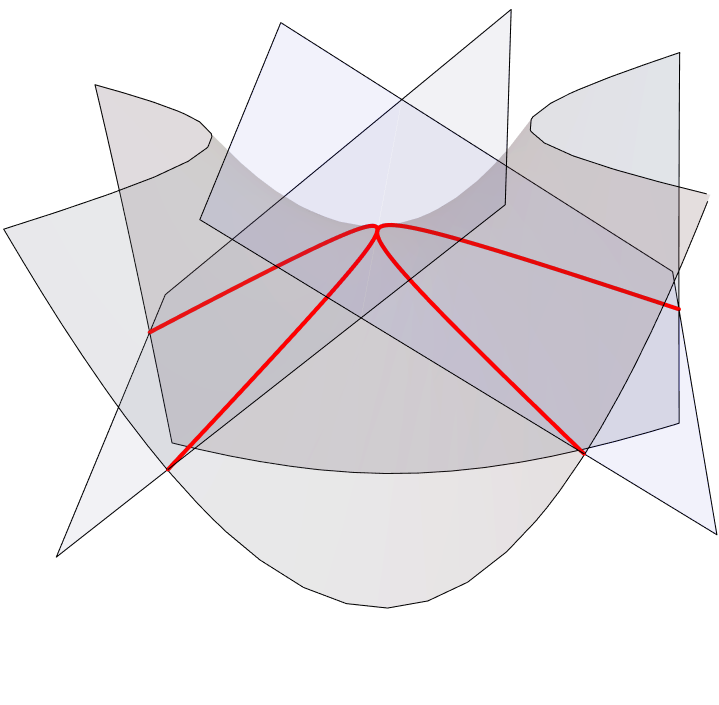}
         \caption{}
     \end{subfigure}
      \hspace{1cm}
     \begin{subfigure}[b]{0.225\textwidth}
         \centering
         \includegraphics[width=0.9\textwidth]{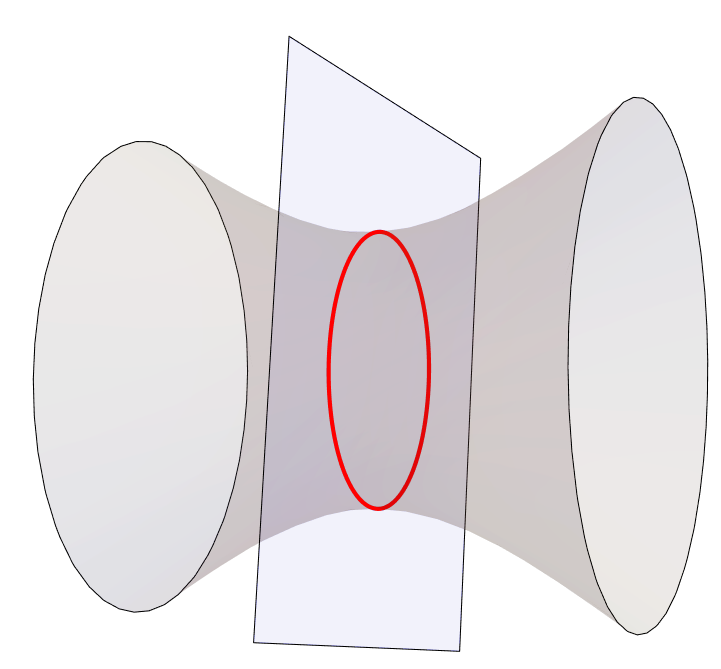}
         \caption{}
     \end{subfigure}\\
          \begin{subfigure}[b]{0.21\textwidth}
         \centering
         \includegraphics[width=\textwidth]{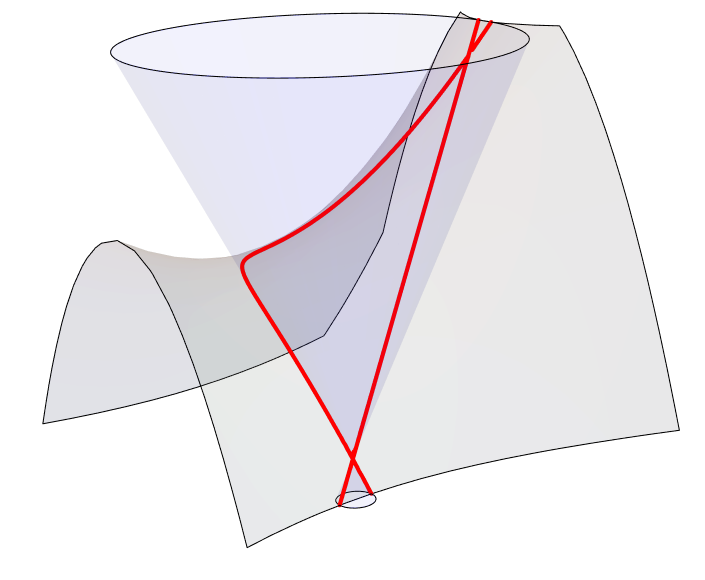}
         \caption{}
     \end{subfigure}
   \hspace{1cm}
     \begin{subfigure}[b]{0.18\textwidth}
         \centering
         \includegraphics[trim={20 30 0 0},width=0.9\textwidth]{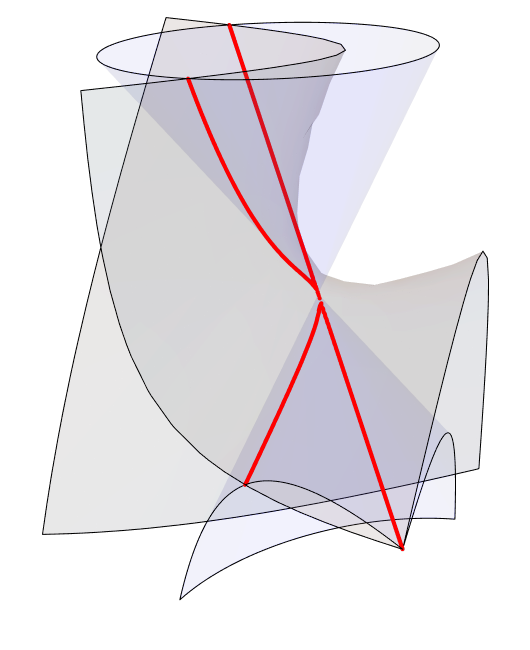}
         \caption{}
     \end{subfigure}
   \hspace{1cm}
     \begin{subfigure}[b]{0.225\textwidth}
         \centering
         \includegraphics[width=0.9\textwidth]{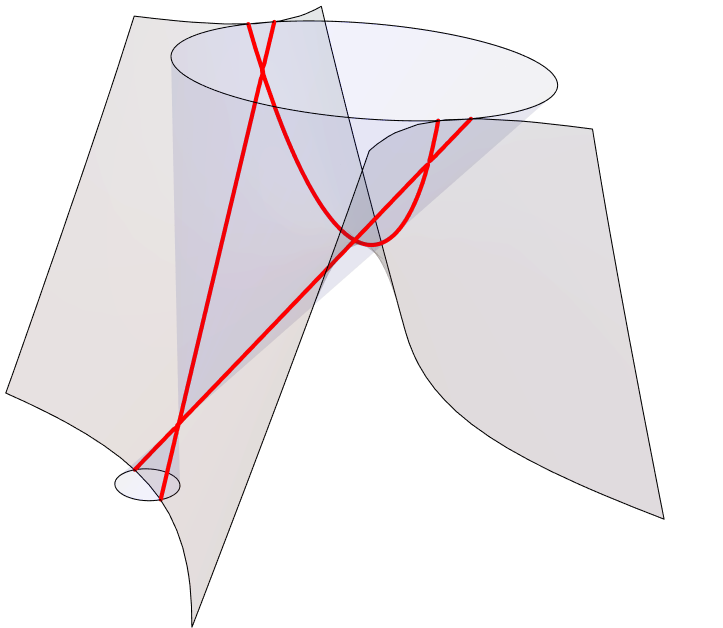}
         \caption{}
     \end{subfigure}\\
          \begin{subfigure}[b]{0.18\textwidth}
         \centering
         \includegraphics[width=\textwidth]{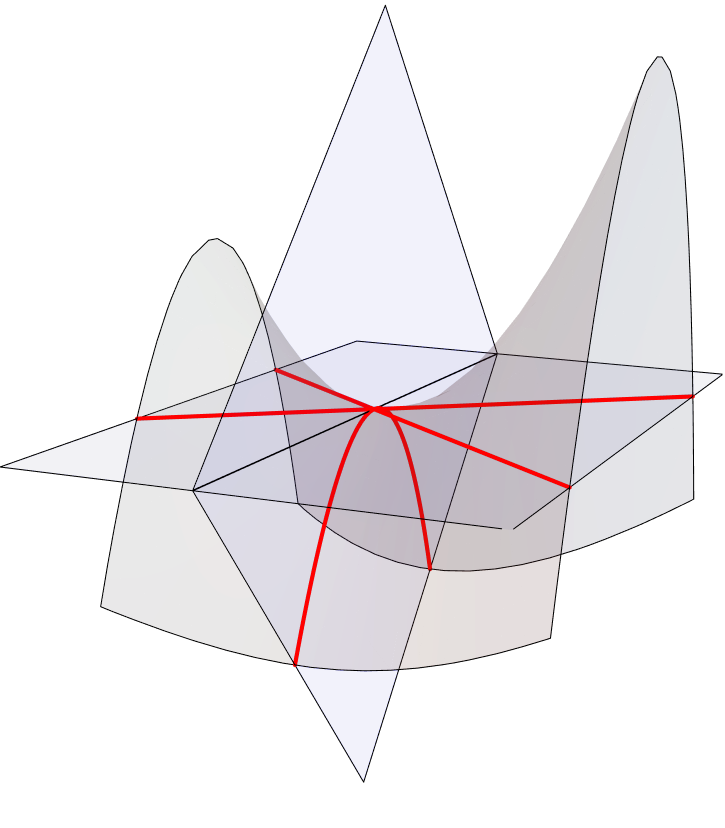}
         \caption{}
     \end{subfigure}
 \hspace{1cm}
     \begin{subfigure}[b]{0.225\textwidth}
         \centering
         \includegraphics[trim={40 40 20 40},width=0.9\textwidth]{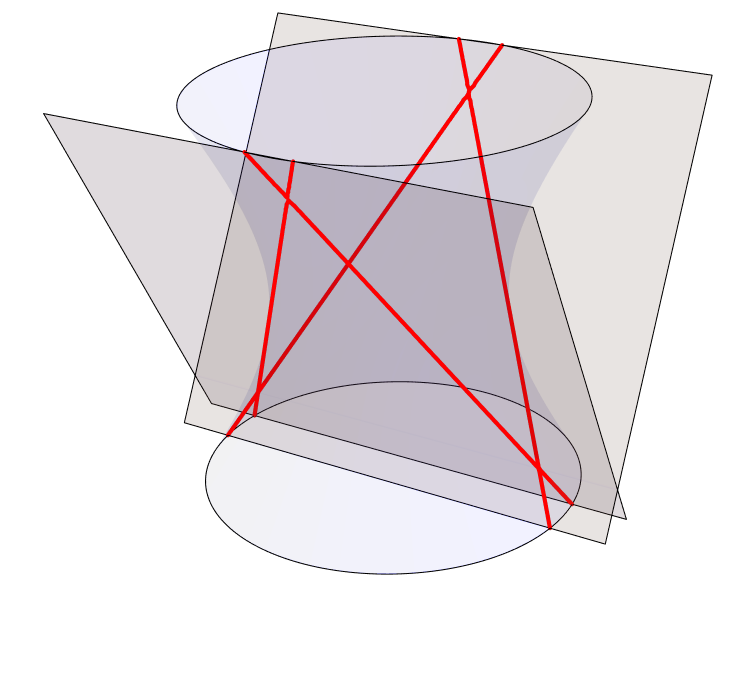}
         \caption{}
     \end{subfigure}
 \hspace{1cm}
     \begin{subfigure}[b]{0.18\textwidth}
         \centering
         \includegraphics[trim={30 0 0 0},width=0.9\textwidth]{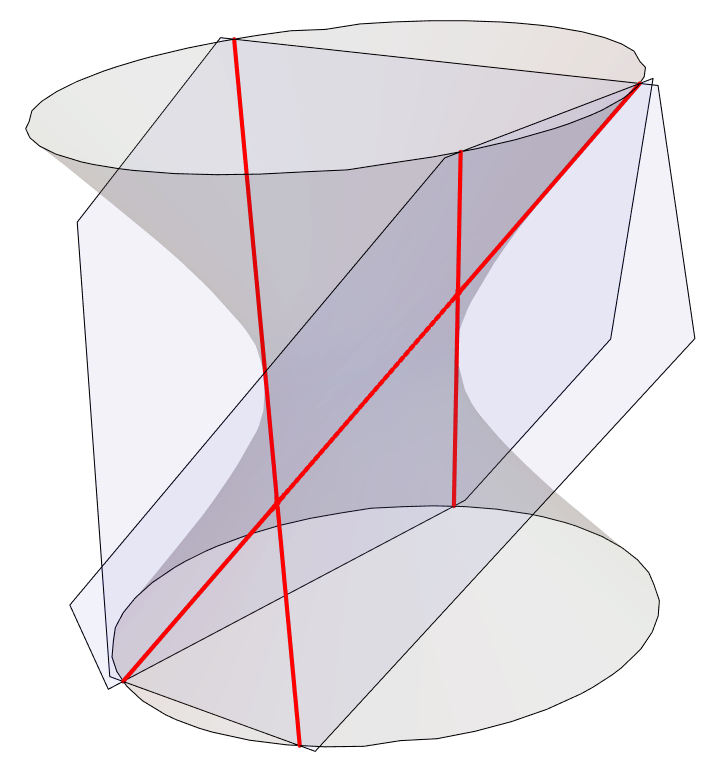}
         \caption{}
      \end{subfigure}
      \hspace{1cm}
     \centering
     \begin{subfigure}[b]{0.18\textwidth}
         \centering
         \includegraphics[trim={0 0 0 -30},width=0.9\textwidth]{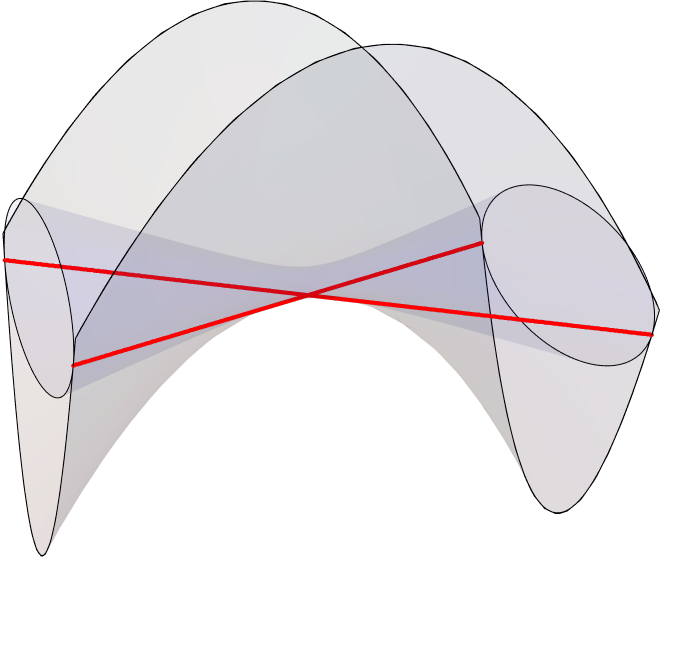}
         \caption{}
     \end{subfigure}
        \caption{The thirteen projective types of pencils of quadrics}
        \label{fig:three graphs}
\end{figure}

Consider the following problem. Suppose that $Q_0(X)=X_1X_2-X_3X_4$. Find a linear projective change of variables $X=A_\lambda Y$ reducing the quadratic form $Q_\lambda$ to the standard form $Q_0$:
\begin{equation}\label{A norm}
Q_\lambda(A_\lambda Y)=Q_0(Y), \quad {\rm or}\quad A_\lambda^{\rm T}M_\lambda A_\lambda=M_0.
\end{equation}
\begin{proposition}
The normalizing matrix $A_\lambda$ is a rational fuction of $\lambda$ and of $\sqrt{\Delta(\lambda)}$. In particular, it is a rational function of $\lambda$ if $\Delta(\lambda)$ is a complete square, i.e., for the seven cases {\rm (vii)--(xiii)}.
\end{proposition}
As a corollary, we obtain what can be called \emph{pencil-adapted coordinates} 
\begin{equation}\label{phi lambda}
\begin{bmatrix} X_1 \\ X_2 \\ X_3 \\ X_4 \end{bmatrix}=A_\lambda\begin{bmatrix} x \\ y \\ xy \\ 1 \end{bmatrix} =:\phi_\lambda(x,y).
\end{equation} 
Thus, $\phi_\lambda$ gives a parametrization of $Q_\lambda$ by $(x,y)\in\bbP^1\times\bbP^1$, such that the generators $\ell_1$, resp. $\ell_2$ of $Q_\lambda$ correspond to $x={\rm const}$, resp. to $y={\rm const}$. 

This establishes a connection to a particular case of a general result of M. Reid \cite[Theorem 1.10]{Reid} on the structure of the set $\ {\rm Gen}(Q)$ of generators of a pencil of quadrics $Q=\{Q_\lambda\}_{\lambda\in\bbP^1}$ in $\bbP^n=\bbP(\bbC^{n+1})$, considered as a subvariety of $\bbP^1\times{\rm Gr}([\frac{n+1}{2}],\bbC^{n+1})$. Reid's theorem says that ${\rm Gen}(Q)$ is a non-singular variety, and for $n$ odd, the first projection ${\rm Gen}(Q)\stackrel{p_1}{\rightarrow} \bbP^1$ factorizes as ${\rm Gen}(Q)\stackrel{p}{\rightarrow} R\stackrel{q}{\rightarrow}\bbP^1$, where $R$ is non-singular, and $q$ is a double covering ramified precisely in ${\rm Sing}(Q)$, and $p$ is smooth. Here, ${\rm Sing}(Q)$ is a finite number of values of $\lambda\in\bbP^1$ for which the quadric $Q_\lambda$ is degenerate.

In practical terms, in our case $n=3$, we can formulate the following statement.
\begin{proposition}
For $X\in Q_\lambda$, the generators $\ell_1(X)$ and $\ell_2(X)$ are rational functions of $X$ and of $\sqrt{\Delta(\lambda)}$. In particular, they are rational functions of $X$ and of $\lambda$ if $\Delta(\lambda)$ is a complete square, i.e., for the seven cases {\rm (vii)--(xiii)}.
\end{proposition}

\section{A 3D QRT map defined by a chosen fiber of the pencil $C_\mu$}
\label{sect 3D QRT for a fiber}

We now specify the construction of a 3D QRT map from Section \ref{sect 3D QRT} by making a special choice of the pencil $Q_\lambda$. 
\medskip

$\bullet\,$ In the pencil of biquadratic curves $\{C_\mu\}$ in $\bbP^1\times\bbP^1$, choose a fiber $C_{\infty}$. We will assume that this is the fiber admitting a decomposition of a canonical type; this condition is not necessary for the construction but will facilitate the discussion of singularity confinement below. Take the corresponding quadric $P_\infty$ in $\bbP^3$, and set
\begin{equation}\label{Qlambda gen}
Q_\lambda=Q_0-\lambda P_\infty.
\end{equation}
Thus, the pencils $\{Q_\lambda\}$ and $\{P_\mu\}$ have one quadric $P_\infty$ in common. The base set of the pencil $\{Q_\lambda\}$ is $Q_0\cap P_\infty=\phi_0(C_\infty)$. Of course, this base set contains the images of the base points $s_1,\ldots,s_8$ of the pencil $\{C_\mu\}$ under the Segre embedding:
\begin{equation}
S_i=\phi_0(s_i), \quad i=1,\ldots,8.
\end{equation}
The set $\{S_1,\ldots,S_8\}$ can be characterized as the intersection of the base curve of the pencil $\{Q_\lambda\}$ with the base curve of the pencil $\{P_\mu\}$, or, alternatively, as the base set of a two-parameter linear family (net) of quadrics spanned by $Q_0$ and $\{P_\mu\}$. Our standing assumption in this paper will be the following. 
\medskip

\noindent
{\bf Assumption.} \emph{The characteristic polynomial $\Delta(\lambda)$ of the pencil $\{Q_\lambda\}$ is a complete square.}
\smallskip

Thus, we will be dealing here with the seven cases (vii)--(xiii) of the classification of pencils considered in Section \ref{sect pencils}. The six cases (i)--(vi) will be dealt with in a follow-up paper \cite{ASWnets}.

$\bullet\,$ Find a linear change of variables $X=A_\lambda Y$ reducing the quadratic form $Q_\lambda(X)$ to the standard form $Q_0(Y)$, as in \eqref{A norm}. The above assumption ensures that $A_\lambda$ is a rational function of $\lambda$. This gives the pencil-adapted coordinates 
$$
\phi_\lambda:\bbP^1\times\bbP^1\to Q_\lambda,
$$ 
as in \eqref{phi lambda}, which can be understood as a parametrization of $Q_\lambda$ by $(x,y)\in\bbP^1\times\bbP^1$, such that the generators $\ell_1$, resp. $\ell_2$ of $Q_\lambda$ correspond to $x={\rm const}$, resp. to $y={\rm const}$.
\medskip

$\bullet\,$  Clearly, for each fixed $\lambda$, the intersection curves $Q_\lambda\cap P_\mu$ in coordinates $(x,y)$ form a pencil of biquadratic curves $\phi_\lambda^*P_\mu$. This pencil can be characterized by its eight base points $s_1(\lambda),\ldots,s_8(\lambda)\in\bbP^1\times\bbP^1$ which are nothing else but the points $S_1,\ldots,S_8$ expressed in the coordinates $\phi_\lambda$:
\begin{equation}\label{s lambda}
\phi_\lambda(s_i(\lambda))=S_i, \quad i=1,\ldots, 8.
\end{equation}
An important object is also the curve $C_\infty(\lambda)\subset\bbP^1\times\bbP^1$ which is just the curve $\phi_0(C_\infty)$ expressed in the coordinates $\phi_\lambda$:
\begin{equation}\label{base curve lambda}
\phi_\lambda(C_\infty(\lambda))=\phi_0(C_\infty).
\end{equation}
Clearly, the curve $C_\infty(\lambda)$ contains the base points $s_1(\lambda),\ldots,s_8(\lambda)$. Moreover, it admits a decomposition of the canonical type, of the same surface type as $C_\infty$.
\medskip

$\bullet\,$  Construct the involutions $i_1,i_2:\bbP^3\dasharrow\bbP^3$ along generators of the pencil $\{Q_\lambda\}$ till the second intersection with $\{P_\mu\}$, and the 3D QRT map $f=i_1\circ i_2:\bbP^3\dasharrow\bbP^3$, as described in Definition \ref{def 3D QRT}. Note that the above assumption ensures that all these maps are birational. One possibility for an effective computing of these maps is to first compute the restrictions $i_1|_{Q_\lambda}$ and $i_2|_{Q_\lambda}$ as the QRT switches corresponding to the pencil $\phi_\lambda^*P_\mu$, and then to push them to homogeneous coordinates $X$ on $\bbP^3$ using \eqref{phi lambda} along with $\lambda=Q_0(X)/P_\infty(X)$. 
\medskip

It is important to observe that the singularity confinement properties of $i_1|_{Q_\lambda}$, $i_2|_{Q_\lambda}$ and of $f|_{Q_\lambda}=i_1|_{Q_\lambda}\circ i_2|_{Q_\lambda}$ are the same as that of the original 2D QRT switches $i_1$, $i_2$ and of the 2D QRT map $f=i_1\circ i_2$, with $s_i=(a_i,b_i)$ being replaced by $s_i(\lambda)=(a_i(\lambda),b_i(\lambda))$. In particular, for $s_i(\lambda)$ not lying on the same vertical or horizontal generator of $\bbP^1\times\bbP^1$ with any other $s_j(\lambda)$, we have a short singularity confinement pattern analogous to \eqref{sing conf QRT i1i2}:
\begin{equation}\label{sing conf 3D QRT i1i2 on fiber}
\{x=a_i(\lambda)\}\; \stackrel{i_1|_{Q_\lambda}}{\longrightarrow}\; s_i(\lambda)\; \stackrel{i_2|_{Q_\lambda}}{\longrightarrow}\; 
\{y=b_i(\lambda)\}.
\end{equation}
For $s_i(\lambda)$, $s_j(\lambda)$ with $a_i(\lambda)=a_j(\lambda)$, we have a long singularity confinement pattern analogous to \eqref{sing conf QRT i1i2 long y}:
\begin{equation}\label{sing conf 3D QRT i1i2 long y on fiber}
\{y=b_i(\lambda)\}\; \stackrel{i_2|_{Q_\lambda}}{\longrightarrow}\; s_i(\lambda) \; \stackrel{i_1|_{Q_\lambda}}{\longrightarrow}\; s_j(\lambda) \; \stackrel{i_2|_{Q_\lambda}}{\longrightarrow}\; \{y=b_j(\lambda)\}.
\end{equation}
And, finally, for $s_i(\lambda)$, $s_j(\lambda)$ with $b_i(\lambda)=b_j(\lambda)$, we have a long singularity confinement pattern analogous to \eqref{sing conf QRT i1i2 long x}:
\begin{equation}\label{sing conf 3D QRT i1i2 long x on fiber}
\{x=a_i(\lambda)\}\; \stackrel{i_1|_{Q_\lambda}}{\longrightarrow}\; s_i(\lambda) \; \stackrel{i_2|_{Q_\lambda}}{\longrightarrow}\; s_j(\lambda) \; \stackrel{i_1|_{Q_\lambda}}{\longrightarrow}\; \{x=a_j(\lambda)\}.
\end{equation}
Now let $\Phi_i\subset\bbP^3$ be the ruled surface consisting of lines on $Q_\lambda$ given, in the pencil-adapted coordinates $\phi_\lambda$, by the equations $\{x=a_i(\lambda)\}$, and let $\Psi_i\subset\bbP^3$ be the ruled surface consisting of lines on $Q_\lambda$ given in the coordinates $\phi_\lambda$ by the equations $\{y=b_i(\lambda)\}$. Then, in view of \eqref{s lambda}, we obtain the following singularity confinement patterns for $i_1, i_2$:
\begin{equation}\label{sing conf 3D QRT i1i2}
\Phi_i\; \stackrel{i_1}{\rightarrow}\; S_i\; \stackrel{i_2}{\rightarrow}\; \Psi_i,
\end{equation}
resp. 
\begin{equation}\label{sing conf 3D QRT i1i2 long y}
\Psi_i\; \stackrel{i_2}{\rightarrow}\; S_i \; \stackrel{i_1}{\rightarrow}\; S_j \; \stackrel{i_2}{\rightarrow}\; \Psi_j
\end{equation}
(if $S_i$ and $S_j$ lie on one $\ell_1$ generator of each $Q_\lambda$), and
\begin{equation}\label{sing conf 3D QRT i1i2 long x}
\Phi_i\; \stackrel{i_1}{\rightarrow}\; S_i \; \stackrel{i_2}{\rightarrow}\; S_j \; \stackrel{i_1}{\rightarrow}\; \Phi_j
\end{equation}
(if $S_i$ and $S_j$ lie on one $\ell_2$ generator of each $Q_\lambda$). The main distinctive feature of these singularity confinement patterns is the blow-down of codimension 1 varieties to points and the blow-up of the points to codimension 1 varieties. This is an ultimate consequence of the fact that the pencils $\{Q_\lambda\}$ and $\{P_\mu\}$ share a common quadric $P_\infty$.


\section{Deforming a 3D QRT map to a 3D Painlev\'e map}
\label{sect L}

\begin{definition}\label{def L}
We call a birational map $L:\bbP^3\dasharrow\bbP^3$ \emph{a Painlev\'e deformation map}, if it satisfies the following conditions:
\begin{itemize}
\item The pencil $\{Q_\lambda\}$ and the base curve $Q_0\cap P_\infty$ are invariant under $L$, but not the individual quadrics $Q_\lambda$. Rather, $L$ maps $Q_\lambda$ to $Q_{\sigma(\lambda)}$, where $\sigma$ is a M\"obius automorphism of $\bbP^1$ fixing the points of $\,{\rm Sing}(Q)=\{\lambda\in\bbP^1: Q_\lambda\;\;{\rm is\;\; singular}\}$. For the cases {\rm (vii), (ix), (xi)} of Section \ref{sect pencils} we have ${\rm Sing}(Q)=\{\lambda_1,\lambda_2\}$, while for the cases {\rm (viii), (x), (xii), (xiii)} we have ${\rm Sing}(Q)=\{\lambda_1\}$. 

\item The singularity confinement properties of $\tilde i_1 :=L\circ i_1$, $\tilde i_2 := L\circ i_2$ are the same as that of $i_1$, $i_2$.
\end{itemize}
Under these conditions, we call $\tilde f := \tilde i_1 \circ \tilde i_2$ \emph{a 3D Painlev\'e map}. 
\end{definition}

The first condition is achieved if we define the action of $L$ on each quadric $Q_\lambda$ individually by
\begin{equation}\label{L on fiber}
L|_{ Q_\lambda}= A_{\sigma(\lambda)}\circ \phi_0\circ \psi_\lambda\circ\phi_0^{-1}\circ A_\lambda^{-1}=\phi_{\sigma(\lambda)}\circ\psi_\lambda\circ \phi_\lambda^{-1}.
\end{equation}
Here, $\psi_\lambda:\bbP^1\times\bbP^1\to\bbP^1\times\bbP^1$ should be chosen to map the curve $C_\infty(\lambda)$ to $C_\infty(\sigma(\lambda))$. In many examples, the curve $C_\infty(\lambda)$ does not depend on $\lambda$, then one can take $\psi_\lambda={\rm id}$, and then
\begin{equation}\label{L on fiber triv}
L|_{Q_\lambda}= A_{\sigma(\lambda)} A_\lambda^{-1}.
\end{equation}
The second condition of Definition \ref{def L}, in principle, has to be verified in each case separately. We formulate here  sufficient conditions which are satisfied in all examples of the present paper.
\begin{proposition}\label{prop Painleve}
\begin{itemize}
\item Suppose that the involutions $i_1, i_2:\bbP^3\dasharrow\bbP^3$ have a singularity confinement pattern of the type \eqref{sing conf 3D QRT i1i2}. If $L$ satisfies
\begin{equation}\label{cond 1}
L(S_i)=S_i,
\end{equation} 
then for the deformed maps $\t i_1=L\circ i_1$, $\t i_2=L\circ i_2$ we have:
\begin{equation}\label{sing conf dP i1i2}
\Phi_i\; \stackrel{\t i_1}{\rightarrow}\; S_i\; \stackrel{\t i_2}{\rightarrow}\; L(\Psi_i),
\end{equation}
which implies for $\t f=\t i_1\circ \t i_2$ the singularity confinement pattern
\begin{equation}\label{sing conf dP f short}
\t i_2^{-1}(\Phi_i)\; \stackrel{\t f}{\rightarrow}\; S_i\; \stackrel{\t f}{\rightarrow}\; (\t i_1\circ L)(\Psi_i).
\end{equation}

\item Suppose that the involutions $i_1, i_2:\bbP^3\dasharrow\bbP^3$ have a singularity confinement pattern of the type \eqref{sing conf 3D QRT i1i2 long y}. If $L$ satisfies
\begin{equation}\label{cond y}
(L\circ i_1\circ L)(S_i)=S_j,
\end{equation} 
then for the deformed maps $\t i_1=L\circ i_1$, $\t i_2=L\circ i_2$ we have:
\begin{equation}\label{sing conf dP i1i2 long y}
\Psi_i\; \stackrel{\t i_2}{\rightarrow}\; L(S_i) \; \stackrel{\t i_1}{\rightarrow}\; S_j \; \stackrel{\t i_2}{\rightarrow}\; L(\Psi_j),
\end{equation}
which implies for $\t f=\t i_1\circ \t i_2$ the singularity confinement pattern
\begin{equation}\label{sing conf dP f long y}
\Psi_i\; \stackrel{\t f}{\rightarrow}\; S_j\; \stackrel{\t f}{\rightarrow}\; (\t i_1\circ L)(\Psi_j).
\end{equation}

\item Suppose that the involutions $i_1, i_2:\bbP^3\dasharrow\bbP^3$ have a singularity confinement pattern of the type \eqref{sing conf 3D QRT i1i2 long x}. If $L$ satisfies
\begin{equation}\label{cond x}
(L\circ i_2\circ L)(S_i)=S_j,
\end{equation} 
then for the deformed maps $\t i_1=L\circ i_1$, $\t i_2=L\circ i_2$ we have:
\begin{equation}\label{sing conf dP i1i2 long x}
\Phi_i\; \stackrel{\t i_1}{\rightarrow}\; L(S_i) \; \stackrel{\t i_2}{\rightarrow}\; S_j \; \stackrel{\t i_1}{\rightarrow}\; L(\Phi_j),
\end{equation}
which implies for $\t f=\t i_1\circ \t i_2$ the singularity confinement pattern
\begin{equation}\label{sing conf dP f long x}
\t i_2^{-1}(\Phi_i)\; \stackrel{\t f}{\rightarrow}\; L(S_i)\; \stackrel{\t f}{\rightarrow}\; L(\Phi_j).
\end{equation}

\end{itemize}
\end{proposition}
\noindent

{\bf Remark.}  In all examples in this paper, the linear system of quadrics through the eight points participating in the  singularity confinement patterns for $\t f$ (these include some of $S_i$ and some of $L(S_i)$) turns out to be  one-dimensional, namely the pencil $Q_\lambda$, if $L\neq{\rm id}$. Of course, if $L={\rm id}$, this linear system is two-dimensional, namely the net based on $S_i$, $i=1,\ldots, 8$, containing both pencils $\{Q_\lambda\}$ and $\{P_\mu\}$.
\smallskip

We now turn to the detailed exposition of the results for all seven types of pencils falling into the framework of the present paper. We start with the pencil of the type (xi) which is the simplest one without infinitely near base points and which provides the reader with the most transparent formulas for all objects involved. Then, we proceed with the types (xii) and (xiii) which involve infinitely near base points and therefore can be considered as degenerations of the type (xi) (moving {\em down} the Sakai's classification). After that, we proceed moving {\em up} the Sakai's classification, which in our terms corresponds to the pencils of the types (x) to (vii) (in this order).

\section{From a pencil of type (xi) to the {\em q}-Painlev\'e equation of the surface type $A_3^{(1)}$}

\begin{figure}[!ht]
\begin{center}
\begin{subfigure}[b]{0.3\textwidth}
         \centering
         \includegraphics[width=0.8\textwidth]{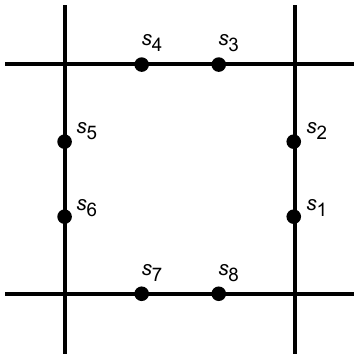}
         \caption{}
     \end{subfigure}
     \hspace{2cm}
     \begin{subfigure}[b]{0.3\textwidth}
         \centering
         \includegraphics[width=\textwidth]{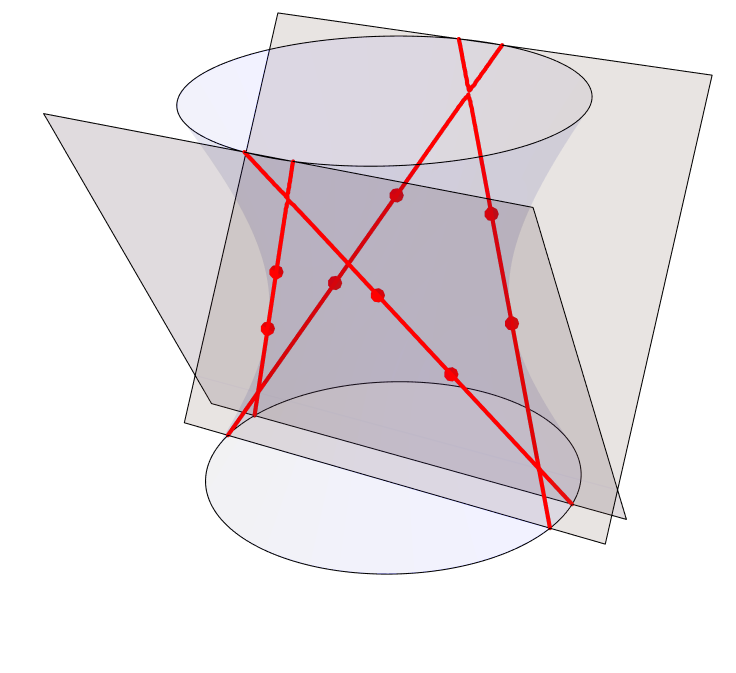}
         \caption{}
     \end{subfigure}
\end{center}
\caption{(a) Base set of the surface type $A_3^{(1)}$: four pairs of points on the sides of a quadrilateral formed by a pair of (1,0)-curves and a pair of (0,1)-curves. (b) Pencil of quadrics through a skew quadrilateral.}
\label{fig qPIII}
\end{figure}

\paragraph{2D QRT map.} We start with the QRT map $f=i_1\circ i_2$ for the pencil of biquadratic curves based on eight points
\begin{align}\label{qPIII QRT p1-p8}
& s_1=(\infty,b_1), \quad s_2=(\infty,b_2), \quad s_3=(a_3,\infty), \quad s_4=(a_4,\infty), \nonumber\\
& s_5=(0,b_5), \quad s_6=(0,b_6), \quad s_7=(a_7,0), \quad s_8=(a_8,0)
\end{align}
A straightforward computation shows that these points support a pencil of biquadratic curves if and only if the following condition is satisfied:
\begin{equation}\label{qPIII QRT condition}
b_1b_2a_7a_8=a_3a_4b_5b_6,
\end{equation}
and then the pencil is given by
\begin{eqnarray}\label{qPIII QRT curves}
C_\mu: && \Big\{\frac{1}{b_1b_2a_7a_8}x^2y^2-\frac{b_1+b_2} {b_1b_2a_7a_8}x^2y-\frac{a_3+a_4} {a_3a_4b_5b_6}x^2y+\frac{1}{a_7a_8}x^2+\frac{1}{b_5b_6}y^2 \nonumber\\
& & \qquad -\frac{a_7+a_8}{a_7a_8}x-\frac{b_5+b_6}{b_5b_6}y+1-\mu xy=0\Big\}.
\end{eqnarray}
The pencil \eqref{qPIII QRT curves} contains a reducible curve, consisting of  a pair of horizontal lines ((0,1)-curves) and a pair ov vertical lines ((1,0)-curves), shown on Fig. \ref{fig qPIII} (a):
\begin{equation}\label{qPIII C infty}
C_\infty=\{xy=0\}: \quad \{x=0\}\cup\{y=0\}\cup\{x=\infty\}\cup\{y=\infty\}.
\end{equation}
\renewcommand\thesubfigure{\alph{subfigure}}

The pencil \eqref{qPIII QRT curves} defines the vertical and the horizontal switches $i_1$, $i_2$:
\begin{equation}\label{qPIII QRT i1}
i_1(x,y)=(x,\t y), \quad {\rm where}\quad \t yy=b_1b_2\frac{(x-a_7)(x-a_8)}{(x-a_3)(x-a_4)}, 
\end{equation}
\begin{equation}\label{qPIII QRT i2}
i_2(x,y)=(\t x,y), \quad {\rm where}\quad \t xx=a_3a_4\frac{(y-b_5)(y-b_6)}{(y-b_1)(y-b_2)},
\end{equation}
and the QRT map 
 \begin{equation} \label{QRT map}
f= i_1\circ i_2.
 \end{equation}
The birational involutions $i_1$, $i_2$ on  $\bbP^1\times \bbP^1$ admit eight ``long'' singularity confinement patterns of the types \eqref{sing conf QRT i1i2 long y}, \eqref{sing conf QRT i1i2 long x} (four of each type). 
From these, eight ``short'' singularity confinement patterns for $f=i_1\circ i_2$ can be easily derived: four of the type \eqref{sing conf QRT f type 2} for $(i,j)=$(1,2), (2,1), (5,6) and (6,5), as well as four of the type \eqref{sing conf QRT f type 3} for $(i,j)=$(3,4), (4,3), (7,8) and (8,7).

\paragraph{3D Painlev\'e map.} We consider the pencil of quadrics $\{P_\mu\}$, the Segre lift of the pencil of curves $\{C_\mu\}$:
\begin{eqnarray}\label{qPIII quadrics}
P_\mu: && \Big\{\frac{1}{b_1b_2a_7a_8}X_3^2-\frac{b_1+b_2} {b_1b_2a_7a_8}X_1X_3-\frac{a_3+a_4} {a_3a_4b_5b_6}X_2X_3+\frac{1}{a_7a_8}X_1^2+\frac{1}{b_5b_6}X_2^2 \nonumber\\
& & \qquad -\frac{a_7+a_8}{a_7a_8}X_1X_4-\frac{b_5+b_6}{b_5b_6}X_2X_4+X_4^2-\mu X_3X_4=0\Big\}.
\end{eqnarray}
We declare the pencil $Q_\lambda$ to be spanned by $Q_0$ and $P_{\infty}=X_3X_4$:
\begin{equation}\label{qPIII pencil}
Q_\lambda=\big\{X_1X_2-\lambda X_3X_4=0\big\}.
\end{equation}
(shift of the parameter $\lambda\to\lambda-1$ is for convenience, to ensure the canonical normalization of ${\rm Sing}(Q)$). The base set of the pencil $Q_\lambda$ is a skew quadrilateral formed by the lines $\{X_1=X_3=0\}$, $\{X_1=X_4=0\}$, $\{X_2=X_3=0\}$, and $\{X_2=X_4=0\}$, see Fig. \ref{fig qPIII} (b). The intersection of this base set with the base set of the pencil $\{P_\mu\}$ consists of eight points
\begin{align}\label{qPIII P1-P8}
& S_1=[1:0:b_1:0], \; S_2=[1:0:b_2:0], \; S_3=[0:1:a_3:0], \; S_4=[0:1:a_4:0], \nonumber\\
& S_5=[0:b_5:0:1], \; S_6=[0:b_6:0:1], \; S_7=[a_7:0:0:1], \; S_8=[a_8:0:0:1],
\end{align}
which are the images of the points $s_1,\ldots,s_8$ given in \eqref{qPIII QRT p1-p8} under the Segre embedding $\phi_0$.

The characteristic polynomial of the pencil $\{Q_\lambda\}$ is $\Delta(\lambda)=\det(M_\lambda)=\lambda^2$, so that ${\rm Sing}(Q)=\{0,\infty\}$.  The 3D QRT involutions $i_1,i_2$ along generators of the pencil $\{Q_\lambda\}$ till the second intersection with $\{P_\mu\}$, as described in Definition \ref{def 3D QRT}, are birational maps of $\mathbb P^3$. They are computed in a straightforward manner and turn out to be of degree 3:
\begin{equation}\label{qPIII i1}
i_1: \begin{bmatrix} X_1 \\ X_2 \\ X_3 \\ X_4 \end{bmatrix} \mapsto 
\begin{bmatrix} \t X_1 \\ \t X_2 \\ \t X_3 \\ \t X_4 \end{bmatrix}=
\begin{bmatrix} X_1(X_3-a_3X_2)(X_3-a_4X_2) \\ b_1b_2X_2 (X_1-a_7X_4)(X_1-a_8X_4)\\
b_1b_2X_3 (X_1-a_7X_4)(X_1-a_8X_4) \\ X_4(X_3-a_3X_2)(X_3-a_4X_2)\end{bmatrix},
\end{equation}
\begin{equation}\label{qPIII i2}
i_2: \begin{bmatrix} X_1 \\ X_2 \\ X_3 \\ X_4 \end{bmatrix} \mapsto 
\begin{bmatrix} \t X_1 \\ \t X_2 \\ \t X_3 \\ \t X_4 \end{bmatrix}=
\begin{bmatrix} a_3a_4X_1 (X_2-b_5X_4)(X_2-b_6X_4)\\
X_2(X_3-b_1X_1)(X_3-b_2X_1) \\ 
a_3a_4X_3 (X_2-b_5X_4)(X_2-b_6X_4) \\ X_4(X_3-b_1X_1)(X_3-b_2X_1)\end{bmatrix}.
\end{equation}

A M\"obius automorphism of $\bbP^1$ fixing ${\rm Sing}(Q)=\{0,\infty\}$ can be taken as $\sigma(\lambda)=q\lambda$ with $q\in\bbC\setminus\{0,1\}$.

\begin{theorem}\label{Th1 for qP(A_3)}
The linear projective map on $\bbP^3$ given by
\begin{equation}\label{qPIII L}
L:\quad X=[X_1:X_2:X_3:X_4]\mapsto [X_1:X_2:q^{-1}X_3:X_4]
\end{equation}
preserves the pencil $\{Q_\lambda\}$ and sends each $Q_\lambda$ to $Q_{q\lambda}$. Moreover, it is a Painlev\'e deformation map: the birational map $\t f=\t i_1\circ \t i_2$ on $\mathbb P^3$ with $\t i_1=L\circ i_1$, $\t i_2=L\circ i_2$ is a 3D Painlev\'e map sending $Q_\lambda$ to $Q_{q^2\lambda}$ with the following singularity confinement patterns:
\begin{itemize}
\item[-] \eqref{sing conf dP f long y} \;for $(i,j)=(1,2), (2,1), (5,6)$ and $(6,5)$, 
\item[-] \eqref{sing conf dP f long x} \;for $(i,j)=(3,4), (4,3), (7,8)$ and $(8,7)$.
\end{itemize}
Here
\begin{align}
& \Psi_1=\{X_3-b_1X_1=0\}, \quad \Psi_2=\{X_3-b_2X_1=0\}, \\
& \Psi_5=\{X_2-b_5X_4=0\}, \quad \Psi_6=\{X_2-b_6X_4=0\}, \\
& \Phi_3=\{X_3-a_3X_2=0\}, \quad \Phi_4=\{X_3-a_4X_2=0\}, \\
& \Phi_7=\{X_1-a_7X_4=0\}, \quad \Phi_8=\{X_1-a_8X_4=0\}.
\end{align}
\end{theorem}
{\bf Proof.} We check by a direct computation that conditions of Proposition \ref{prop Painleve} are satisfied. Namely:
\begin{itemize}
\item $L$ fixes the points $S_i$, $i=5,6,7,8$. This ensures that \eqref{cond y} is satisfied for $(i,j)=(5,6), (6,5)$, and  \eqref{cond x} is satisfied for  $(i,j)=(7,8), (8,7)$;
\item $L\circ i_1$ maps $L(S_1)$ to $S_2$ and $L(S_2)$ to $S_1$, so that \eqref{cond y} is satisfied for $(i,j)=(1,2), (2,1)$;
\item $L\circ i_2$ maps $L(S_3)$ to $S_4$ and $L(S_4)$ to $S_3$, so that \eqref{cond x} is satisfied for  $(i,j)=(3,4), (4,3)$.
\end{itemize}
We mention also that  $L$ fixes the planes $\Psi_i=\{X_2=b_iX_4\}$ for $i=5,6$, and the planes $\Phi_i=\{X_1=a_iX_4\}$ for $i=7,8$, maps the planes  $\Psi_i=\{X_3=b_iX_1\}$ to the planes $L(\Psi_i)=\{qX_3=b_iX_1\}$ for $i=1,2$, and the planes $\Phi_i=\{X_3=a_iX_2\}$ to the planes $L(\Phi_i)=\{qX_3=a_iX_2\}$ for $i=3,4$. $\;\blacksquare$
\smallskip

\noindent
{\bf Remark.} The eight points participating in the singularity confinement patterns for $\t f$ are: $S_i$ for $i=1,2$ and $i=5,6,7,8$, and $L(S_i)$ for $i=3,4$. If $q\neq \pm 1$, then the linear system of quadrics through these eight points is one-dimensional, namely the pencil $Q_\lambda$.  If $q=\pm 1$, it is a two-dimensional net spanned by the pencils $\{Q_\lambda\}$ and $\{P_\mu\}$.

\paragraph{Relation to the $q$-Painlev\'e equation of the surface type $A_3^{(1)}$.} To establish a relation between the map $\t f$ and a $q$-Painlev\'e equation, we start by computing the normalizing transformation of $Q_\lambda$ to the canonical form $Q_0$: 
\begin{equation}\label{qPIII pencil norm}
 \begin{bmatrix} X_1 \\ X_2\\ X_3\\ X_4\end{bmatrix}=
 \begin{bmatrix} Y_1 \\ Y_2\\ \lambda^{-1}Y_3\\ Y_4\end{bmatrix}
 =A_\lambda\begin{bmatrix} Y_1\\Y_2\\Y_3\\Y_4\end{bmatrix}, \quad
A_\lambda=\begin{pmatrix} 1 & 0 & 0 &  0\\ 0 & 1 & 0 & 0 \\ 0 & 0 & \lambda^{-1} & 0 \\ 0 & 0 & 0 & 1\end{pmatrix}.                               
 \end{equation}
 This immediately gives the following parametrization of $Q_\lambda$: 
\begin{equation}\label{qPIII pencil x to X}
 \begin{bmatrix} X_1 \\ X_2\\ X_3\\ X_4\end{bmatrix}=A_\lambda\begin{bmatrix} x \\ y\\  xy\\   1\end{bmatrix}
 =\begin{bmatrix} x \\ y\\  \lambda^{-1}xy\\   1\end{bmatrix}=:\phi_\lambda(x,y).
 \end{equation}
Thus, the pencil-adapted coordinates $(x,y,\lambda)$ on $\bbP^3$ are given by
 \begin{equation}\label{qPIII pencil X to x}
 x=\frac{X_1}{X_4}=\lambda\frac{X_3}{X_2}, \qquad y=\frac{X_2}{X_4}=\lambda\frac{X_3}{X_1}, \qquad \lambda=\frac{X_1X_2}{X_3X_4}.
 \end{equation}

In the pencil-adapted coordinates $(x,y,\lambda)$, for each fixed $\lambda$, the intersection curves $Q_\lambda\cap P_\mu$ form the pencil $\phi_\lambda^* P_\mu$ which can be characterized as the pencil of biquadratic curves in $\bbP^1\times\bbP^1$ through the eight points
\begin{align}\label{qPIII p1-p8 lambda}
& s_1(\lambda)=(\infty,b_1\lambda), \quad s_2(\lambda)=(\infty,b_2\lambda), \nonumber\\
& s_3(\lambda)=(a_3\lambda,\infty), \quad s_4(\lambda)=(a_4\lambda,\infty),\nonumber\\
& s_5(\lambda)=(0,b_5), \quad s_6(\lambda)=(0,b_6), \quad s_7(\lambda)=(a_7,0), \quad s_8(\lambda)=(a_8,0),
\end{align}
which correspond to $S_1,\ldots,S_8$ given in \eqref{qPIII P1-P8} under the map $\phi_\lambda^{-1}$. The curve $C_\infty(\lambda)$ has the same equation $\{xy=0\}$ as the curve $C_\infty$ and is given by \eqref{qPIII C infty}.
Pencil $\phi_\lambda^* P_\mu$ can be obtained from \eqref{qPIII QRT curves} by the modification of parameters $b_i\mapsto b_i\lambda$, $i=1,2$, and $a_i\mapsto a_i\lambda$, $i=3,4$. Therefore, formulas for the involutions $i_1$, $i_2$ restricted to $Q_\lambda$  coincide with the original formulas \eqref{qPIII QRT i1}, \eqref{qPIII QRT i2}, with the modified parameters:
\begin{equation}\label{qPIII 3D i1 on fiber}
i_1|_{Q_\lambda}:\; (x,y)\mapsto (x,\t y), \;\; {\rm where}\;\; \t yy=\frac{b_1b_2}{a_3a_4}\cdot \frac{(x-a_7)(x-a_8)}{\big(1-(a_3\lambda)^{-1}x\big)
\big(1-(a_4\lambda)^{-1}x\big)},
\end{equation}
\begin{equation}\label{qPIII 3D i2 on fiber}
i_2|_{Q_\lambda}: \; (x,y)\mapsto (\t x,y), \;\; {\rm where} \;\; \t xx=\frac{a_3a_4}{b_1b_2}\cdot \frac{(y-b_5)(y-b_6)}{\big(1-(b_1\lambda)^{-1}y\big)\big(1-(b_2\lambda)^{-1}y\big)}.
\end{equation}
There follows:
\begin{theorem}
If one parametrizes $Q_\lambda$ by $(x,y)\in\bbP^1\times \bbP^1$ according to \eqref{qPIII pencil x to X}, then in coordinates $(x,y,\lambda)$ on $\bbP^3$ the map $\t f:(x_n,y_n,\lambda_{2n})\mapsto(x_{n+1},y_{n+1},\lambda_{2n+2})$ is equivalent to the $q$-Painlev\'e equation of the surface type $A_3^{(1)}$, a system of two non-autonomous difference equations:
\begin{eqnarray}
x_{n+1}x_n & = & \frac{a_3a_4}{b_1b_2}\cdot \frac{(y_n-b_5)(y_n-b_6)}{\big(1-(b_1\lambda_{2n})^{-1}y_n\big)\big(1-(b_2\lambda_{2n})^{-1}y_n\big)}, \label{qPIII x} \\
y_{n+1}y_n & = & \frac{b_1b_2}{a_3a_4}\cdot \frac{(x_{n+1}-a_7)(x_{n+1}-a_8)}{\big(1-(a_3\lambda_{2n+1})^{-1}x_{n+1}\big)\big(1-(a_4\lambda_{2n+1})^{-1}x_{n+1}\big)}, \label{qPIII y}
\end{eqnarray}
where $\lambda_n=q^n\lambda_0$.
\end{theorem}

\noindent
{\bf Computational remarks.} The map $L$ given in \eqref{qPIII L} can be found as follows: since the curves $C_\infty(\lambda)$ on $\bbP^1\times\bbP^1$ do not depend on $\lambda$, we can take $L|_{Q_\lambda}=A_{\sigma(\lambda)}A_\lambda^{-1}$. A simple computation confirms that the right-hand side does not depend on $\lambda$, and is given by \eqref{qPIII L}. Likewise, the pencil adapted coordinates are handy to compute the surfaces $\Psi_i$ and $\Phi_i$. Indeed, the maps $i_1|_{Q_\lambda}$, $i_2|_{Q_\lambda}$ admit eight ``long'' singularity confinement patterns of the types \eqref{sing conf 3D QRT i1i2 long y on fiber}, \eqref{sing conf 3D QRT i1i2 long x on fiber} (four of each type). They are easily translated to homogeneous coordinates, becoming \eqref{sing conf 3D QRT i1i2 long y} for $(i,j)=(1,2) , (2,1), (5,6)$ and $(6,5)$, and
\eqref{sing conf 3D QRT i1i2 long x} for $(i,j)=(3,4), (4,3), (7,8)$ and $(8,7)$, with the equations for $\Psi_i$ and $\Phi_i$ computed with the help of \eqref{qPIII pencil X to x}. Similar remarks hold true for the other examples, as well.


\section{From a pencil of type (xii) to the {\em d}-Painlev\'e equation of the surface type $D_4^{(1)}$ }
\label{sect D4}

\paragraph{2D QRT map.} We start with the QRT map corresponding to a pencil of biquadratic curves in $\bbP^1\times\bbP^1$ through the following eight points:
\begin{equation}
s_1=(0,b_1),\quad s_2=(0,b_2),\quad s_3=(\infty,b_3), \quad s_4=(\infty,b_4),
\end{equation}
\begin{equation}
s_5=(a_5,\infty),\quad s_6=(a_5(1+a_6\epsilon),\epsilon^{-1}),\quad s_7=(a_7,\infty), \quad s_8=(a_7(1+a_8\epsilon),\epsilon^{-1}).
\end{equation}
Here, $s_6$ and $s_8$ are infinitely close points to $s_5$ and $s_7$, respectively, and the $\epsilon$ notation means that if we plug in the expressions of $s_6$ and $s_8$ into the equation of the curve, then the resulting expression vanishes up to the first order in $\epsilon$. 

\begin{figure}[!ht]
\begin{center}
\begin{subfigure}[b]{0.3\textwidth}
         \centering
         \includegraphics[width=0.8\textwidth]{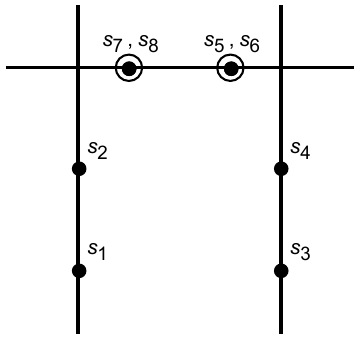}
         \caption{}
     \end{subfigure}
     \hspace{2cm}
     \begin{subfigure}[b]{0.3\textwidth}
         \centering
         \includegraphics[width=\textwidth]{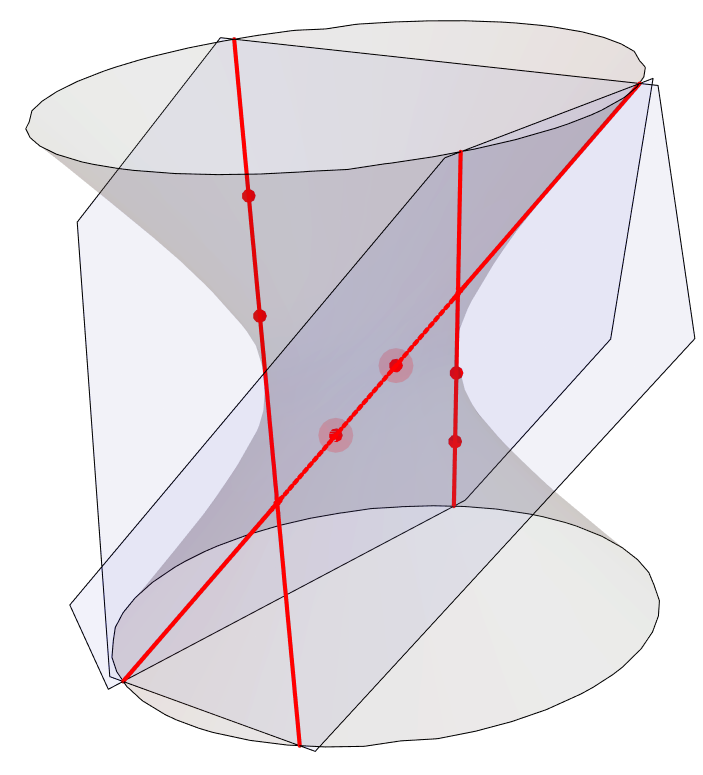}
         \caption{}
     \end{subfigure}
\end{center}
\caption{(a) Base set of the surface type $D_4^{(1)}$: two pairs of points on two (1,0)-curves and two pairs of infinitely near points on one double (0,1)-curve. (b) Pencil of quadrics through two lines and their common transversal line, tangent along the latter.}
\label{fig dP D4}
\end{figure}

One easily computes that such eight points are base points of a pencil of biquadratic curves if and only if
$$
a_6+a_8=b_3+b_4-b_1-b_2.
$$
The pencil contains a reducible curve $C_\infty$ with the equation $\{x=0\}$, which consists of the following irreducible components:
\begin{equation}\label{D4 C infty}
C_\infty=\{x=0\}\cup \{x=\infty\}\cup \{y=\infty\}^2
\end{equation}
(the last component counts as a double line), see Fig. \ref{fig dP D4} (a).

The vertical switch $i_1$ and the horizontal switch $i_2$ for this pencil are given by the following formulas:
\begin{equation}\label{QRT D4 i1}
i_1(x,y)=(x,\t y), \quad {\rm where}\quad \t y+y=b_3+b_4+\frac{a_5a_6}{x-a_5}+\frac{a_7a_8}{x-a_7},
\end{equation}
\begin{equation}\label{QRT D4 i2}
i_2(x,y)=(\t x,y), \quad {\rm where}\quad\t xx=a_5a_7\frac{(y-b_1)(y-b_2)}{(y-b_3)(y-b_4)}.
\end{equation}
The maps $i_1$, $i_2$ have four ``long'' singularity confinement patterns of the type \eqref{sing conf QRT i1i2 long y} (for $(i,j)=$(1,2), (2,1), (3,4) and (4,3), as well as  two ``long'' singularity confinement patterns of the type \eqref{sing conf QRT i1i2 long x},
\begin{equation} \label{dP D4 QRT sing 3}
\{x=a_5\}\;\;\stackrel{i_1}{\to}\;\:s_6\;\;\stackrel{i_2}{\to}\;\:s_8 \;\;\stackrel{i_1}{\to} \;\;\{x=a_7\},
\end{equation}
and the similar one with the roles of $s_6$ and $s_8$ exchanged.

\paragraph{3D Painlev\'e map.} We consider the pencil of quadrics $\{P_\mu\}$, the Segre lift of the pencil of curves $\{C_\mu\}$, and we declare the pencil $Q_\lambda$ to be spanned by $Q_0$ and $P_{\infty}=X_1X_4$:
\begin{equation}\label{D4 pencil}
Q_\lambda=\{X_1X_2-X_3X_4-\lambda X_1X_4=0\}.
\end{equation}
Its base curve consists of two lines $\{X_1=X_3=0\}$, $\{X_2=X_4=0\}$, and a double line $\{X_1=X_4=0\}$, see Fig. \ref{fig dP D4} (b). The intersection of this base set with the base set of the pencil $\{P_\mu\}$ consists of eight points
\begin{align}\label{D4 P1-P8}
& S_1=[0:b_1: 0:1], \; \; S_2=[0:b_2:0: 1], \;\;  S_3=[1:0:b_3:0],\;\;  S_4=[1:0:b_4:0], \nonumber\\
& S_5=[0:1:a_5:0], \;\;  S_6=[a_5\epsilon:1:a_5(1+a_6\epsilon):\epsilon],\nonumber\\ 
& S_7=[0:1:a_7:0], \;\;  S_8=[a_7\epsilon:1:a_7(1+a_8\epsilon):\epsilon],
\end{align}
where $S_6$ and $S_8$ are understood as infinitely near points to $S_5$ and $S_7$, respectively.
The characteristic polynomial of the pencil $\{Q_\lambda\}$ equals $\Delta(\lambda)=\det(M_\lambda)=1$, so that ${\rm Sing}(Q)=\{\infty\}$. The 3D QRT involutions $i_1,i_2$ along generators of the pencil $\{Q_\lambda\}$ till the second intersection with $\{P_\mu\}$, and the 3D QRT map $f=i_1\circ i_2$, as described in Definition \ref{def 3D QRT}, are birational maps of $\mathbb P^3$. While the formulas for the involution $i_2$ in homogeneous coordinates are relatively simple:
\begin{equation}\label{D4 3D QRT i2}
i_2: \begin{bmatrix} X_1 \\ X_2 \\ X_3 \\ X_4 \end{bmatrix} \mapsto 
\begin{bmatrix} \t X_1 \\ \t X_2 \\ \t X_3 \\ \t X_4 \end{bmatrix}=
\begin{bmatrix} a_5a_7X_1 (X_2-b_1X_4)(X_2-b_2X_4)\\
X_2(X_3-b_3X_1)(X_3-b_4X_1) \\ 
a_5a_7X_3 (X_2-b_1X_4)(X_2-b_2X_4) \\ X_4(X_3-b_3X_1)(X_3-b_4X_1)\end{bmatrix},
\end{equation}
the formulas for $i_1$ are somewhat messy: $i_1:[X_1:X_2:X_3:X_4]\mapsto[\t X_1:\t X_2:\t X_3:\t X_4]$, where
\begin{eqnarray}
\t X_1 & = & X_1(X_1-a_5X_4)(X_1-a_7X_4), \nonumber\\
\t X_2 & = & X_1^2X_2-2X_1X_3X_4+(b_3+b_4)X_1^2X_4+(a_5+a_7)X_3X_4^2-a_5a_7X_2X_4^2\nonumber\\
 & & +\big(a_5a_6+a_7a_8-(a_5+a_7)(b_3+b_4)\big)X_1X_4^2+a_5a_7(b_1+b_2)X_4^3,\nonumber\\
 \t X_3 & = & -X_1^2X_3+(b_3+b_4)X_1^3+(a_5+a_7)X_1^2X_2+a_5a_7X_3X_4^2-2a_5a_7X_1X_2X_4\nonumber\\
 & & +\big(a_5a_6+a_7a_8-(a_5+a_7)(b_3+b_4)\big)X_1^2X_4+a_5a_7(b_1+b_2)X_1X_4^2,\nonumber\\
 \t X_4 & = & X_4(X_1-a_5X_4)(X_1-a_7X_4). \label{D4 3D QRT i1}
\end{eqnarray}

A M\"obius automorphism of $\bbP^1$ fixing ${\rm Sing}(Q)=\{\infty\}$ can be taken as 
$\sigma(\lambda)=\lambda+\delta$ with $\delta\in\bbC\setminus\{0\}$.

\begin{theorem}\label{Th1 for dP(D_4)}
The linear projective map on $\bbP^3$ given by 
\begin{equation}\label{D4 L}
L:\quad [X_1:X_2:X_3:X_4]\to [X_1:X_2:X_3-\delta X_1:X_4]
\end{equation}
preserves the pencil $\{Q_\lambda\}$ and sends each $Q_\lambda$ to $Q_{\lambda+\delta}$. Moreover, it is a Painlev\'e deformation map: the birational map $\t f=\t i_1\circ \t i_2$ on $\mathbb P^3$ with $\t i_1=L\circ i_1$, $\t i_2=L\circ i_2$ is a 3D Painlev\'e map sending $Q_\lambda$ to $Q_{\lambda+2\delta}$ with the following singularity confinement patterns:
\begin{itemize}
\item[-] \;\eqref{sing conf dP f long y}\; for $(i,j)=(1,2), (2,1), (3,4), (4,3)$, 
\item[-]  \;\eqref{sing conf dP f long x}\; for $(i,j)=(6,8), (8,6)$.
\end{itemize}
Here
$$
\Psi_i=\{X_2-b_i X_4=0\}, \;i=1,2, \quad \Psi_i=\{X_3-b_i X_1=0\}, \;i=3,4, 
$$
$$
\Phi_6=\{X_1-a_5 X_4=0\}, \quad \Phi_8=\{X_1-a_7 X_4=0\}.
$$
\end{theorem}
{\bf Proof.} This follows by a slight adaption of the arguments of Proposition \ref{prop Painleve} (taking into account the infinitely near points). Namely: 
\begin{itemize}
\item Map $L$ fixes the points $S_1$, $S_2$, $S_5$, $S_7$, while 
$$
L(S_3)=[1:0:b_3-\delta:0],\quad L(S_4)=[1:0:b_4-\delta:0],
$$
$$ 
L(S_6)=[a_5\epsilon:1:a_5(1+(a_6-\delta)\epsilon):\epsilon],\quad L(S_8)=[a_7\epsilon:1:a_7(1+(a_8-\delta)\epsilon):\epsilon].
$$
\item $L\circ i_1$ maps $L(S_3)$ to $S_4$ and $L(S_4)$ to $S_3$;
\item $L\circ i_2$ maps $L(S_6)$ to $S_8$ and $L(S_8)$ to $S_6$.
\end{itemize}
All this follows by a direct computation. $\blacksquare$
\medskip

{\bf Remark.}  The eight points participating in the singularity confinement patterns for $\widetilde f$ are: the six points $S_1,S_2,S_3,S_4,S_5,S_7$ and the two infinitely near points $L(S_6), L(S_8)$. If $\delta\neq 0$, they support a one-dimensional linear system of quadrics, namely the pencil $Q_\lambda$. If $\delta=0$, this set becomes the two-dimensional net spanned by $\{Q_\lambda\}$ and $\{P_\mu\}$.

\paragraph{Relation to the $d$-Painlev\'e equation of the surface type $D_4^{(1)}$.} To establish a relation between the map $\t f$ and a $d$-Painlev\'e equation, we start by computing the normalizing transformation of $Q_\lambda$ to the canonical form $Q_0$: 
\begin{equation}\label{D4 pencil norm}
 \begin{bmatrix} X_1 \\ X_2\\ X_3\\ X_4\end{bmatrix}=
 \begin{bmatrix} Y_1 \\ Y_2\\ Y_3-\lambda Y_1\\ Y_4\end{bmatrix}
 =A_\lambda\begin{bmatrix} Y_1\\Y_2\\Y_3\\Y_4\end{bmatrix}, \quad
A_\lambda=\begin{pmatrix} 1 & 0 & 0 &  0\\ 0 & 1 & 0 & 0 \\ -\lambda & 0 & 1 & 0 \\ 0  & 0 & 0 & 1\end{pmatrix}.                               
 \end{equation}
This gives the following parametrization of $Q_\lambda$: 
\begin{equation}\label{D4 pencil x to X}
 \begin{bmatrix} X_1 \\ X_2\\ X_3\\ X_4\end{bmatrix}=A_\lambda\begin{bmatrix} x \\ y\\  xy\\   1\end{bmatrix}
 =\begin{bmatrix} x \\ y \\  xy-\lambda x\\ 1 \end{bmatrix}=:\phi_\lambda(x,y).
 \end{equation}
The pencil-adapted coordinates $(x,y,\lambda)$ on  $\bbP^3$ are:
\begin{equation}\label{D4 pencil X to x}
  x=\frac{X_1}{X_4}=\frac{X_3+\lambda X_1}{X_2}, \quad
  y=\frac{X_2}{X_4}=\frac{X_3+\lambda X_1}{X_1}, \quad
   \lambda=\frac{X_1X_2-X_3X_4}{X_1X_4}.
\end{equation}
In the pencil-adapted coordinates $(x,y,\lambda)$, for each fixed $\lambda$, the intersection curves $Q_\lambda\cap P_\mu$ form the pencil $\phi_\lambda^* P_\mu$ which can be characterized as the pencil of biquadratic curves in $\bbP^1\times\bbP^1$ through the eight points
\begin{align}\label{D4 p1-p8 lambda}
& s_1(\lambda)=(0,b_1), \quad s_2(\lambda)=(0,b_2),\quad s_3(\lambda)=(\infty,b_3+\lambda),\quad s_4(\lambda)=(\infty,b_4+\lambda). \nonumber\\
& s_5(\lambda)=(a_5,\infty),\quad s_6=(a_5(1+(a_6+\lambda)\epsilon),\epsilon^{-1}),\nonumber\\
& s_7(\lambda)=(a_7,\infty),\quad s_8=(a_7(1+(a_8+\lambda)\epsilon),\epsilon^{-1}),
\end{align}
which correspond to $S_1,\ldots,S_8$ given in \eqref{D4 P1-P8} under the map $\phi_\lambda^{-1}$. The curve $C_\infty(\lambda)$ coincides with $C_\infty$. 
Formulas for the involutions $i_1$, $i_2$ restricted to $Q_\lambda$ are obtained from \eqref{QRT D4 i1}, \eqref{QRT D4 i2} by replacing $b_i\mapsto b_i+\lambda$  for $i=3,4$, and $a_i\mapsto a_i+\lambda$  for $i=6,8$: 
\begin{equation}\label{D4 3D QRT i1 on fiber}
i_1|_{Q_\lambda}(x,y)=(x,\t y), \quad {\rm where}\quad \t y+y=b_3+b_4+2\lambda+\frac{a_5(a_6+\lambda)}{x-a_5}+\frac{a_7(a_8+\lambda)}{x-a_7},
\end{equation}
\begin{equation}\label{D4 3D QRT i2 on fiber}
i_2|_{Q_\lambda}(x,y)=(\t x,y), \quad {\rm where}\quad\t xx=a_5a_7\frac{(y-b_1)(y-b_2)}{(y-b_3-\lambda)(y-b_4-\lambda)}.
\end{equation}

\begin{theorem}
If one parametrizes $Q_\lambda$ by $(x,y)\in\bbP^1\times \bbP^1$ according to \eqref{D4 pencil x to X}, then in coordinates $(x,y,\lambda)$ on $\bbP^3$ the map $\t f:(x_n,y_n,\lambda_{2n})\mapsto(x_{n+1},y_{n+1},\lambda_{2n+2})$ is equivalent to the $d$-Painlev\'e equation of the surface type $D_4^{(1)}$, a system of two non-autonomous difference equations:
\begin{eqnarray}
x_{n+1}x_n & = & a_5a_7\frac{(y_n-b_1)(y_n-b_2)}{(y_n-b_3-\lambda_{2n})(y_n-b_4-\lambda_{2n})},
 \label{dP D4 x} \\
y_{n+1}+y_n & = & b_3+b_4+2\lambda_{2n+1}+\frac{a_5(a_6+\lambda_{2n+1})}{x_{n+1}-a_5}+\frac{a_7(a_8+\lambda_{2n+1})}{x_{n+1}-a_7},
\label{dP D4 y}
\end{eqnarray}
where $\lambda_n=n\delta+\lambda_0$.
\end{theorem}

\section{From a pencil of type (xiii) to the {\em d}-Painlev\'e equation of the type $D_5^{(1)}$ }

\begin{figure}[h!]
\begin{center}
\begin{subfigure}[b]{0.3\textwidth}
         \centering
         \includegraphics[width=0.8\textwidth]{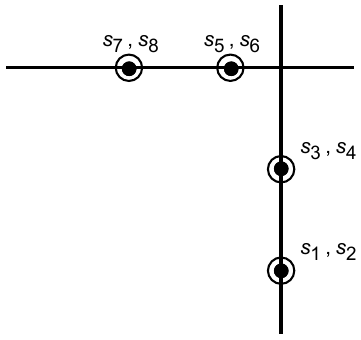}
         \caption{}
     \end{subfigure}
     \hspace{2cm}
     \begin{subfigure}[b]{0.3\textwidth}
         \centering
         \includegraphics[width=\textwidth]{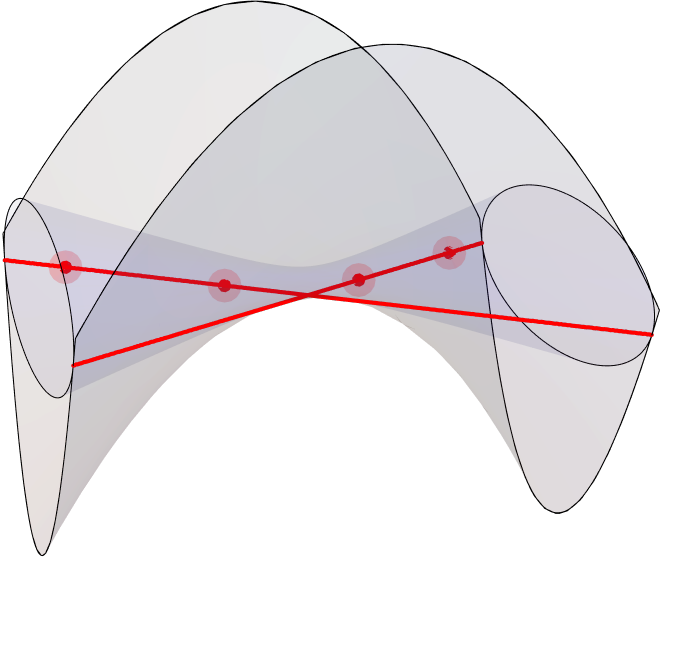}
         \caption{}
     \end{subfigure}
\end{center}
\caption{(a) Base set of the surface type $D_5^{(1)}$: four pairs of infinitely near points on a double (0,1)-curve and a double (1,0)-curve. (b) Pencil of quadrics tangent along a pair of intersecting lines.}
\label{fig dPII}
\end{figure}

\paragraph{2D QRT map.} We start with a QRT map corresponding to the pencil of biquadratic curves in $\bbP^1\times\bbP^1$ through the following eight points: four finite points
\begin{equation}
s_1=(\infty,b_1), \quad s_3=(\infty,b_3), \quad s_5=(a_5,\infty), \quad s_7=(a_7,\infty),
\end{equation}
and four further infinitely near points
\begin{align}
& s_2=(\epsilon^{-1},b_1+b_2\epsilon), \quad s_4=(\epsilon^{-1},b_3+b_4\epsilon), \\
& s_6=(a_5+a_6\epsilon,\epsilon^{-1}), \quad s_8=(a_7+a_8\epsilon,\epsilon^{-1}).
\end{align}
A direct computation shows that these points form a base set for a biquadratic pencil if and only if the following condition is satisfied:
\begin{equation}\label{dPII condition}
b_2+b_4=a_6+a_8.
\end{equation}
The pencil contains a reducible curve $C_\infty$ corresponding to a biquadratic polynomial $1$. This curve consists of two double lines:
\begin{equation}\label{dPII C infty}
C_\infty=\{x=\infty\}^2\cup\{y=\infty\}^2,
\end{equation}
see Fig. \ref{fig dPII} (a).

The vertical switch $i_1$ and the horizontal switch $i_2$ for this pencil are given by the following formulas:
\begin{equation}\label{dPII QRT i1}
i_1(x,y)=(x,\t y), \quad {\rm where}\quad \t y+y=b_1+b_3+\frac{a_6}{x-a_5}+\frac{a_8}{x-a_7},
\end{equation}
\begin{equation}\label{dPII QRT i2}
i_2(x,y)=(\t x,y), \quad {\rm where}\quad \t x+x=a_5+a_7+\frac{b_2}{y-b_1}+\frac{b_4}{y-b_3}.
\end{equation}
The maps $i_1$, $i_2$ have four ``long'' singularity confinement patterns:
\begin{align} 
& \{y=b_1\}\;\;\stackrel{i_2}{\to}\;\:s_2\;\;\stackrel{i_1}{\to}\;\:s_4 \;\;\stackrel{i_2}{\to} \;\;\{y=b_3\},
\label{dPII QRT sing 1}\\
& \{y=b_3\}\;\;\stackrel{i_2}{\to}\;\:s_4\;\;\stackrel{i_1}{\to}\;\:s_2 \;\;\stackrel{i_2}{\to} \;\;\{y=b_1\},
\label{dPII QRT sing 2}\\
& \{x=a_5\}\;\;\stackrel{i_1}{\to}\;\:s_6\;\;\stackrel{i_2}{\to}\;\:s_8 \;\;\stackrel{i_1}{\to} \;\;\{x=a_7\},
\label{dPII QRT sing 3}\\
& \{x=a_7\}\;\;\stackrel{i_1}{\to}\;\:s_8\;\;\stackrel{i_2}{\to}\;\:s_6 \;\;\stackrel{i_1}{\to} \;\;\{x=a_5\}.
\label{dPII QRT sing 4}
\end{align}
We give here the ``naive'' singularity confinement patterns. This means that we display blow-down of $\mathcal C(i_1)$, $\mathcal C(i_2)$ under $i_1$, resp. $i_2$; thus, we do not perform the ``last'' blow-ups which regularize the lifts of these maps.

\paragraph{3D Painlev\'e map.} We consider the pencil of quadrics $\{P_\mu\}$, the Segre lift of the pencil of curves $\{C_\mu\}$, and we declare the pencil $Q_\lambda$ to be spanned by $Q_0$ and $P_{\infty}=X_4^2$:
\begin{equation}\label{dPII pencil}
Q_\lambda=\big\{X_1X_2-X_3X_4-\lambda X_4^2=0\big\}.
\end{equation}
The base set of the pencil $Q_\lambda$ consists of two double lines $\{X_1=X_4=0\}$ and $\{X_2=X_4=0\}$, see Fig. \ref{fig dPII} (b). The intersection of this base set with the base set of the pencil $\{P_\mu\}$ consists of eight points
\begin{align}\label{dPII P1-P8}
& S_1=[1:0:b_1: 0], \; \; S_3=[1:0:b_3:0], \;\;  S_5=[0:1:a_5:0],\;\;  S_7=[0:1:a_7:0], \nonumber\\
& S_2=[1:b_1\epsilon:b_1+b_2\epsilon:\epsilon], \;\;  S_4=[1:b_3\epsilon:b_3+b_4\epsilon:\epsilon], \nonumber\\ 
& S_6=[a_5\epsilon:1:a_5+a_6\epsilon:\epsilon], \;\;  S_8=[a_7\epsilon:1:a_7+a_8\epsilon:\epsilon],
\end{align}
where $S_2,S_4,S_6,S_8$ are understood as infinitely near points to $S_1,S_3,S_5,S_7$, respectively.
The characteristic polynomial of the pencil $\{Q_\lambda\}$ equals $\Delta(\lambda)=\det(M_\lambda)=1$, so that ${\rm Sing}(Q)=\{\infty\}$. The 3D QRT involutions $i_1,i_2$ along generators of the pencil $\{Q_\lambda\}$ till the second intersection with $\{P_\mu\}$, and the 3D QRT map $f=i_1\circ i_2$, as described in Definition \ref{def 3D QRT}, are birational maps of $\mathbb P^3$. The involutions $i_1,i_2$ are of $\deg=3$ and given by formulas similar to \eqref{D4 3D QRT i1}. For instance, $i_1:[X_1:X_2:X_3:X_4]\mapsto[\t X_1:\t X_2:\t X_3:\t X_4]$, where
\begin{eqnarray}
\t X_1 & = & X_1(X_1-a_5X_4)(X_1-a_7X_4), \nonumber\\
\t X_2 & = & X_1^2X_2-2X_1X_3X_4+(b_1+b_3)X_1^2X_4+(a_5+a_7)X_3X_4^2-a_5a_7X_2X_4^2 \nonumber\\
 & & +\big(a_6+a_8-(b_1+b_3)(a_5+a_7)\big)X_1X_4^2- \big(a_6a_7+a_5a_8-a_5a_7(b_1+b_3)\big)X_4^3, \nonumber\\
 \t X_3 & = & -X_1^2X_3+(b_1+b_3)X_1^3+(a_5+a_7)X_1^2X_2+a_5a_7X_3X_4^2-2a_5a_7X_1X_2X_4 \nonumber\\
 & & +\big(a_6+a_8-(b_1+b_3)(a_5+a_7)\big)X_1^2X_4- \big(a_6a_7+a_5a_8-a_5a_7(b_1+b_3)\big)X_1X_4^2, \nonumber\\
 \t X_4 & = & X_4(X_1-a_5X_4)(X_1-a_7X_4).  \label{D5 3D QRT i1}
\end{eqnarray} 

A M\"obius automorphism of $\bbP^1$ fixing ${\rm Sing}(Q)=\{\infty\}$ can be taken as 
$\sigma(\lambda)=\lambda+\delta$ with $\delta\in\bbC\setminus\{0\}$.

\begin{theorem}\label{Th1 for dP(D_5)}
The linear projective map on $\bbP^3$ given by 
\begin{equation}\label{dPII L}
L:\quad [X_1:X_2:X_3:X_4]\to [X_1:X_2:X_3-\delta X_4:X_4].
\end{equation}
preserves the pencil $\{Q_\lambda\}$ and sends each $Q_\lambda$ to $Q_{\lambda+\delta}$. Moreover, it is a Painlev\'e deformation map: the birational map $\t f=\t i_1\circ \t i_2$ on $\mathbb P^3$ with $\t i_1=L\circ i_1$, $\t i_2=L\circ i_2$ is a 3D Painlev\'e map sending $Q_\lambda$ to $Q_{\lambda+2\delta}$ with the following singularity confinement patterns:
\begin{itemize}
\item[-] \;\eqref{sing conf dP f long y}\; for $(i,j)=(2,4), (4,2)$, 
\item[-]  \;\eqref{sing conf dP f long x}\; for $(i,j)=(6,8), (8,6)$.
\end{itemize}
Here
\begin{eqnarray*}
\Psi_2=\{X_2-b_1 X_4=0\}, && \Psi_4=\{X_2-b_3 X_4=0\}, \\
\Phi_6=\{X_1-a_5 X_4=0\}, && \Phi_8=\{X_1-a_7 X_4=0\}.
\end{eqnarray*}
\end{theorem}
{\bf Proof.} Follows from Proposition \ref{prop Painleve} by observing that:
\begin{itemize}
\item The map $L$ fixes the points $S_1,S_3,S_5,S_7$, while it maps the infinitely near points as follows:
$$
L(S_2)=[1:b_1\epsilon:b_1+(b_2-\delta)\epsilon:\epsilon], \quad L(S_4)=[1:b_3\epsilon:b_3+(b_4-\delta)\epsilon:\epsilon], 
$$
$$ 
L(S_6)=[a_5\epsilon:1:a_5+(a_6-\delta)\epsilon:\epsilon],\quad L(S_8)=[a_7\epsilon:1:a_7+(a_8-\delta)\epsilon:\epsilon];
$$
\item $L\circ i_1$ maps $L(S_2)$ to $S_4$ and $L(S_4)$ to $S_2$;
\item $L\circ i_2$ maps $L(S_6)$ to $S_8$ and $L(S_8)$ to $S_6$.
\end{itemize}
We notice also that $L$ fixes the four planes $\Psi_2$, $\Psi_4$, $\Phi_6$, $\Phi_6$. $\blacksquare$
\medskip

{\bf Remark.}  The eight points participating in the singularity confinement patterns for $\widetilde f$ are: the four points $S_1,S_3,S_5,S_7$ and the four infinitely near points $S_2, S_4, L(S_6), L(S_8)$. If $\delta\neq 0$, they support a one-dimensional linear system of quadrics, namely the pencil $Q_\lambda$. If $\delta=0$, this set becomes the two-dimensional net spanned by $\{Q_\lambda\}$ and $\{P_\mu\}$.

\paragraph{Relation to the $d$-Painlev\'e equation of the surface type $D_5^{(1)}$.} To establish a relation between the map $\t f$ and a $d$-Painlev\'e equation, we start by computing the normalizing transformation of $Q_\lambda$ to the canonical form $Q_0$: 
\begin{equation}\label{dPII pencil norm}
 \begin{bmatrix} X_1 \\ X_2\\ X_3\\ X_4\end{bmatrix}=
 \begin{bmatrix} Y_1 \\ Y_2\\ Y_3-\lambda Y_4\\ Y_4\end{bmatrix}
 =A_\lambda\begin{bmatrix} Y_1\\Y_2\\Y_3\\Y_4\end{bmatrix}, \quad
A_\lambda=\begin{pmatrix} 1 & 0 & 0 &  0\\ 0 & 1 & 0 & 0 \\ 0 & 0 & 1 & -\lambda \\ 0  & 0 & 0 & 1\end{pmatrix}.                               
 \end{equation}
This gives the following parametrization of $Q_\lambda$: 
\begin{equation}\label{dPII pencil x to X}
 \begin{bmatrix} X_1 \\ X_2\\ X_3\\ X_4\end{bmatrix}=A_\lambda\begin{bmatrix} x \\ y\\  xy\\   1\end{bmatrix}
 =\begin{bmatrix} x \\ y \\  xy-\lambda \\ 1 \end{bmatrix}=:\phi_\lambda(x,y).
 \end{equation}
The pencil-adapted coordinates $(x,y,\lambda)$ on  $\bbP^3$ are:
\begin{equation}\label{dPII pencil X to x}
  x=\frac{X_1}{X_4}=\frac{X_3+\lambda X_4}{X_2}, \quad
  y=\frac{X_2}{X_4}=\frac{X_3+\lambda X_4}{X_1}, \quad
   \lambda=\frac{X_1X_2-X_3X_4}{X_4^2}.
\end{equation}
In the pencil-adapted coordinates $(x,y,\lambda)$, for each fixed $\lambda$, the intersection curves $Q_\lambda\cap P_\mu$ form the pencil $\phi_\lambda^* P_\mu$ which can be characterized as the pencil of biquadratic curves in $\bbP^1\times\bbP^1$ through the eight points
\begin{align}\label{dPII p1-p8 lambda}
& s_1(\lambda)=(\infty,b_1), \quad s_3(\lambda)=(\infty,b_3),\quad s_5(\lambda)=(a_5,\infty),\quad s_7(\lambda)=(a_7,\infty), \nonumber\\
& s_2(\lambda)=(\epsilon^{-1},b_1+(b_2+\lambda)\epsilon),\quad s_4(\lambda)=(\epsilon^{-1},b_3+(b_4+\lambda)\epsilon),
\nonumber\\
& s_6(\lambda)=(a_5+(a_6+\lambda)\epsilon,\epsilon^{-1}),\quad  s_8(\lambda)=(a_7+(a_8+\lambda)\epsilon,\epsilon^{-1}).
\end{align}
The latter points correspond to $S_1,\ldots,S_8$ given in \eqref{dPII P1-P8} under the map $\phi_\lambda^{-1}$. The curve $C_\infty(\lambda)$ coincides with $C_\infty$. 
Formulas for the involutions $i_1$, $i_2$ restricted to $Q_\lambda$ are obtained from \eqref{dPII QRT i1}, \eqref{dPII QRT i2} by replacing $b_i\mapsto b_i+\lambda$  for $i=2,4$, and $a_i\mapsto a_i+\lambda$  for $i=6,8$: 
\begin{equation}\label{dPII 3D QRT i1 on fiber}
i_1|_{Q_\lambda}(x,y)=(x,\t y), \quad {\rm where}\quad  \t y+y=b_1+b_3+\frac{a_6+\lambda}{x-a_5}+\frac{a_8+\lambda}{x-a_7},\end{equation}
\begin{equation}\label{dPII 3D QRT i2 on fiber}
i_2|_{Q_\lambda}(x,y)=(\t x,y), \quad {\rm where}\quad \t x+x=a_5+a_7+\frac{b_2+\lambda}{y-b_1}+\frac{b_4+\lambda}{y-b_3}.
\end{equation}

\begin{theorem}
If one parametrizes $Q_\lambda$ by $(x,y)\in\bbP^1\times \bbP^1$ according to \eqref{D4 pencil x to X}, then in coordinates $(x,y,\lambda)$ on $\bbP^3$ the map $\t f:(x_n,y_n,\lambda_{2n})\mapsto(x_{n+1},y_{n+1},\lambda_{2n+2})$ is equivalent to the additive Painlev\'e equation of the type $D_5^{(1)}$\!, a system of two non-autonomous difference equations:
\begin{eqnarray}
x_{n+1}+x_n & = & a_5+a_7+\frac{b_2+\lambda_{2n}}{y_n-b_1}+\frac{b_4+\lambda_{2n}}{y_n-b_3},
 \label{dPII x} \\
y_{n+1}+y_n & = & b_1+b_3+\frac{a_6+\lambda_{2n+1}}{x_{n+1}-a_5}+\frac{a_8+\lambda_{2n+1}}{x_{n+1}-a_7},
\label{dPII y}
\end{eqnarray}
where $\lambda_n=n\delta+\lambda_0$.
\end{theorem}

\section{From a pencil of type (x)  to {\em d}-Painlev\'e equation of the surface type $A_2^{(1)}$ }

\begin{figure}[!ht]
\begin{center}
\begin{subfigure}[b]{0.3\textwidth}
         \centering
         \includegraphics[width=0.8\textwidth]{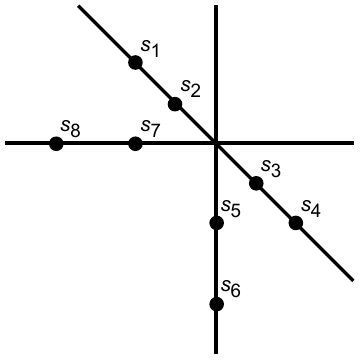}
         \caption{}
     \end{subfigure}
     \hspace{2cm}
     \begin{subfigure}[b]{0.3\textwidth}
         \centering
         \includegraphics[width=\textwidth]{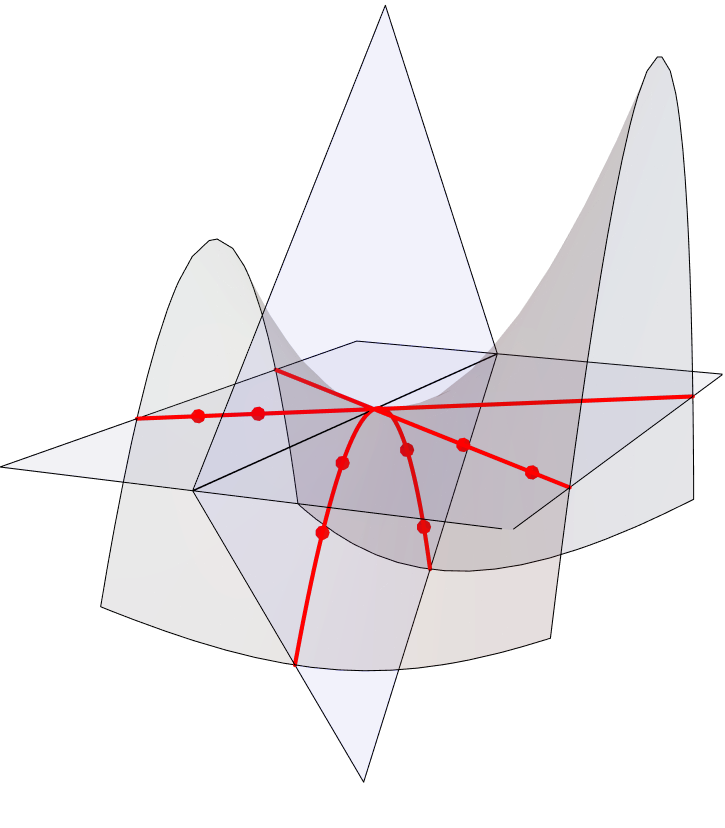}
         \caption{}
     \end{subfigure}
\end{center}
\caption{(a) Base set of the surface type $A_2^{(1)}$: four points on a (1,1)-curve and two pairs of points on two generators intersecting on the curve. (b) Pencil of quadrics through a conic and two lines meeting on the conic.}
\label{fig dPIV}
\end{figure}

\paragraph{2D QRT map.} We start with the QRT map $f=i_1\circ i_2$ for the pencil of biquadratic curves based on eight points
\begin{equation}
s_i=(a_i,-a_i), \quad i=1,\ldots,4,
\end{equation}
\begin{equation}
s_5=(\infty,b_5), \quad s_6=(\infty,b_6), \quad s_7=(a_7,\infty), \quad s_8=(a_8,\infty).
\end{equation}
A straightforward computation shows that these points support a pencil of biquadratic curves if and only if the following condition is satisfied:
\begin{equation}\label{dPIV QRT condition}
a_1+a_2+a_3+a_4+b_5+b_6=a_7+a_8.
\end{equation}
The pencil through these eight points contains a reducible curve $C_\infty$ with the equation $x+y=0$, consisting of the following three irreducible components: 
\begin{equation}\label{dPIV C infty}
C_\infty: \quad\{x=\infty\}\cup\{y=\infty\}\cup\{x+y=0\}.
\end{equation}
This curve is shown on Fig. \ref{fig dPIV} (a). 

The vertical and the horizontal switches $i_1$, $i_2$ for the above mentioned pencil are:
\begin{equation}\label{dPIV QRT i1}
i_1(x,y)=(x,\t y), \quad {\rm where}\quad (\t y+x)(y+x)=\frac{\prod_{i=1}^4(x-a_i)}{(x-a_7)(x-a_8)}, 
\end{equation}
\begin{equation}\label{dPIV QRT i2}
i_2(x,y)=(\t x,y), \quad {\rm where}\quad (\t x+y)(x+y)=\frac{\prod_{i=1}^4(y+a_i)}{(y-b_5)(y-b_6)},
\end{equation}
and the corresponding QRT map is $f= i_1\circ i_2$.
The birational involutions $i_1$, $i_2$ on  $\bbP^1\times \bbP^1$ admit four ``short'' singularity confinement patterns \eqref{sing conf QRT i1i2} and four ``long'' singularity confinement patterns of the types \eqref{sing conf QRT i1i2 long y}, \eqref{sing conf QRT i1i2 long x} (two of each type). 

\paragraph{3D Painlev\'e map.} We consider the pencil of quadrics $\{P_\mu\}$, the Segre lift of the pencil of curves $\{C_\mu\}$, and we declare the pencil $Q_\lambda$ to be spanned by $Q_0$ and $P_{\infty}=(X_1+X_2)X_4$:
\begin{equation}\label{dPIV pencil}
Q_\lambda=\Big\{X_1X_2-X_3X_4-\lambda(X_1+X_2)X_4=0\Big\}.
\end{equation}
The base set of the pencil $Q_\lambda$ consists of the the conic $\{X_1X_2-X_3X_4=0,\; X_1+X_2=0\}$, and two lines $\{X_1=X_4=0\}$, $\{X_2=X_4=0\}$ intersecting in the point $[0:0:1:0]$ on the conic, see Fig. \ref{fig dPIV} (b). Intersection of this base set with the base set of the pencil $\{P_\mu\}$ consists of eight points 
\begin{align}\label{dPIV P1-P8}
& S_i=[a_i:-a_i:-a_i^2:1], \quad i=1,\ldots4, \nonumber\\
& S_5=[1:0:b_5:0], \quad S_6=[1:0:b_6:0], \nonumber\\ 
& S_7=[0:1:a_7:0], \quad S_8=[0:1:a_8:0].
\end{align}
The characteristic polynomial of the pencil $\{Q_\lambda\}$ equals $\Delta(\lambda)=\det(M_\lambda)=1$, so that ${\rm Sing}(Q)=\{\infty\}$. The 3D QRT involutions $i_1,i_2$ along generators of the pencil $\{Q_\lambda\}$ till the second intersection with $\{P_\mu\}$, and the 3D QRT map $f=i_1\circ i_2$, as described in Definition \ref{def 3D QRT}, are birational maps of $\mathbb P^3$. 

A M\"obius automorphism of $\bbP^1$ fixing ${\rm Sing}(Q)=\{\infty\}$ can be taken as 
$\sigma(\lambda)=\lambda+\delta$ with $\delta\in\bbC\setminus\{0\}$.
\begin{theorem}\label{Th1 for dP(A_2)}
The linear projective map on $\bbP^3$ given by
\begin{equation}\label{dPIV L}
L:\quad X=[X_1:X_2:X_3:X_4]\mapsto [X_1:X_2:X_3-\delta(X_1+X_2):X_4].
\end{equation}
preserves the pencil $\{Q_\lambda\}$ and sends each $Q_\lambda$ to $Q_{\lambda+\delta}$. Moreover, it is a Painlev\'e deformation map: the birational map $\t f=\t i_1\circ \t i_2$ on $\mathbb P^3$ with $\t i_1=L\circ i_1$, $\t i_2=L\circ i_2$ is a 3D Painlev\'e map sending $Q_\lambda$ to $Q_{\lambda+2\delta}$ with the following singularity confinement patterns:
\begin{itemize}
\item[-] \;\eqref{sing conf dP f short}\; for $i=1,\ldots,4$, 
\item[-] \;\eqref{sing conf dP f long y}\; for $(i,j)=(5,6), (6,5)$, 
\item[-]  \;\eqref{sing conf dP f long x}\; for $(i,j)=(7,8), (8,7)$.
\end{itemize}
Here
\begin{align}
& \Phi_i=\{X_1-a_iX_4=0\}, \quad \Psi_i=\{X_2+a_iX_4=0\}, \quad i=1,\ldots,4,\\
& \Psi_i=\{X_2^2+X_3X_4-b_iX_4(X_1+X_2)=0\}, \quad i=5,6 \\
& \Phi_i=\{X_1^2+X_3X_4-a_iX_4(X_1+X_2)=0\}, \quad i=7,8.
\end{align}
\end{theorem}
{\bf Proof.} All these statements are demonstrated by a direct computation. In particular, to prove the statement about singularity confinement, one checks that conditions of Proposition \ref{prop Painleve} are satisfied:
\begin{itemize}
\item $L$ fixes the points $S_i$, $i=1,\ldots,4$, and acts on the other four base points as follows:
\begin{align}
& L(S_5)=[1:0:b_5-\delta:0], \quad L(S_6)=[1:0:b_6-\delta:0], \\
& L(S_7)=[0:1:a_7-\delta:0], \quad L(S_8)=[0:1:a_8-\delta:0];
\end{align}
\item $L\circ i_1$ maps $L(S_5)$ to $S_6$ and $L(S_6)$ to $S_5$;
\item $L\circ i_2$ maps $L(S_7)$ to $S_8$ and $L(S_8)$ to $S_7$.
\end{itemize}
We notice also that $L$ fixes the planes $\Psi_i$ and $\Phi_i$ for $i=1,\ldots,4$, maps the quadrics  $\Psi_i$ to analogous quadrics with $b_i\mapsto b_i-\delta$ for $i=5,6$, and maps the quadrics $\Phi_i$ to analogous quadrics with $a_i\mapsto a_i-\delta$ for $i=7,8$. $\blacksquare$
\medskip

\noindent
{\bf Remark.}  The eight points participating in the above singularity confinement patterns for $\t f$ are: $S_i$, $i=1,\ldots,6$, and $L(S_7)$, $L(S_8)$. If $\delta\neq 0$, then the linear system of quadrics through these eight points is one-dimensional, namely the pencil $Q_\lambda$.  If $\delta=0$, it is two-dimensional, namely the net containing both pencils $\{Q_\lambda\}$ and $\{P_\mu\}$. Similar remark holds true for the examples in the next four sections, as well.

\paragraph{Relation to the $d$-Painlev\'e equation of the surface type $A_2^{(1)}$.} To establish a relation between the map $\t f$ and a $d$-Painlev\'e equation, we start by computing the normalizing transformation of $Q_\lambda$ to the canonical form $Q_0$: 
\begin{equation}\label{dPIV pencil norm}
 \begin{bmatrix} X_1 \\ X_2\\ X_3\\ X_4\end{bmatrix}=
 \begin{bmatrix} Y_1 \\ Y_2\\ Y_3-\lambda(Y_1+Y_2)\\ Y_4\end{bmatrix}
 =A_\lambda\begin{bmatrix} Y_1\\Y_2\\Y_3\\Y_4\end{bmatrix}, \quad
A_\lambda=\begin{pmatrix} 1 & 0 & 0 &  0\\ 0 & 1 & 0 & 0 \\ -\lambda & -\lambda & 1 & 0 \\ 0 & 0 & 0 & 1\end{pmatrix}.                               
 \end{equation}
 This immediately gives the following parametrization of $Q_\lambda$: 
\begin{equation}\label{dPIV pencil x to X}
\begin{bmatrix} X_1 \\ X_2 \\ X_3 \\ X_4 \end{bmatrix}
=\begin{bmatrix} x \\ y \\ xy-\lambda(x+y) \\ 1  \end{bmatrix}=:\phi_\lambda(x,y).
\end{equation}
The pencil-adapted coordinates $(x,y,\lambda)$ on $\bbP^3$ are:
\begin{equation}\label{dPIV pencil X to x}
x = \frac{X_1}{X_4}=\frac{X_3+\lambda X_2}{X_2-\lambda X_4}, \quad y = \frac{X_2}{X_4}=\frac{X_3+\lambda X_1}{X_1-\lambda X_4}, \quad \lambda =\frac{X_1X_2-X_3X_4}{(X_1+X_2)X_4}.
\end{equation}

In the pencil-adapted coordinates $(x,y,\lambda)$, for each fixed $\lambda$, the intersection curves $Q_\lambda\cap P_\mu$ form the pencil $\phi_\lambda^* P_\mu$ which can be characterized as the pencil of biquadratic curves in $\bbP^1\times\bbP^1$ through the eight points
\begin{align}\label{dPIV p1-p8 lambda}
& s_i(\lambda)=(a_i,-a_i), \quad i=1,\ldots,4, \nonumber\\
& s_5(\lambda)=(\infty,b_5+\lambda), \quad s_6(\lambda)=(\infty,b_6+\lambda), \nonumber\\
& s_7(\lambda)=(a_7+\lambda,\infty), \quad s_8(\lambda)=(a_8+\lambda,\infty),
\end{align}
which correspond to $S_1,\ldots,S_8$ given in \eqref{dPIV P1-P8} under the map $\phi_\lambda^{-1}$. The curve $C_\infty(\lambda)$ has the same equation $\{x+y=0\}$ as the curve $C_\infty$ and is given by \eqref{dPIV C infty}.
Pencil $\phi_\lambda^* P_\mu$ can be obtained from $C_\mu$ by the modification of parameters $b_i\mapsto b_i+\lambda$, $i=5,6$, and $a_i\mapsto a_i+\lambda$, $i=7,8$. Therefore, formulas for the involutions $i_1$, $i_2$ restricted to $Q_\lambda$  coincide with the original formulas \eqref{dPIV QRT i1}, \eqref{dPIV QRT i2}, with the modified parameters:
\begin{align}\label{dPIV 3D i1 on fiber}
i_1|_{Q_\lambda}:\; (x,y)\mapsto (x,\t y), \; {\rm where}\;\;(\t y+x)(y+x)=\frac{\prod_{i=1}^4(x-a_i)}{(x-a_7-\lambda)(x-a_8-\lambda)},\\
\label{dPIV 3D i2 on fiber}
i_2|_{Q_\lambda}: \; (x,y)\mapsto (\t x,y), \; {\rm where} \;\;  (\t x+y)(x+y)=\frac{\prod_{i=1}^4(y+a_i)}{(y-b_5-\lambda)(y-b_6-\lambda)}.
\end{align}
There follows:
\begin{theorem}
If one parametrizes $Q_\lambda$ by $(x,y)\in\bbP^1\times \bbP^1$ according to \eqref{dPIV pencil x to X}, then in coordinates $(x,y,\lambda)$ on $\bbP^3$ the map $f_\delta:(x_n,y_n,\lambda_{2n})\mapsto(x_{n+1},y_{n+1},\lambda_{2n+2})$ is equivalent to the $d$-Painlev\'e equation of the surface type $A_2^{(1)}$, a system of two non-autonomous difference equations:
\begin{eqnarray}
 (x_{n+1}+y_n)(x_n+y_n) & = & \frac{\prod_{i=1}^4(y_n+a_i)}{(y_n-b_5-\lambda_{2n})(y_n-b_6-\lambda_{2n})},
 \label{dPIV x} \\
(y_{n+1}+x_{n+1})(y_n+x_{n+1}) & = & \frac{\prod_{i=1}^4(x_{n+1}-a_i)}{(x_{n+1}-a_7-\lambda_{2n+1})(x_{n+1}-a_8-\lambda_{2n+1})},
 \label{dPIV y}
\end{eqnarray}
where $\lambda_n=n\delta+\lambda_0$.
\end{theorem}

\section{From a pencil of type (ix) to the {\em q}-Painlev\'e equation of the surface type $A_2^{(1)}$ }
\label{sect case ix}

\begin{figure}[!ht]
\begin{center}
\begin{subfigure}[b]{0.3\textwidth}
         \centering
         \includegraphics[width=0.8\textwidth]{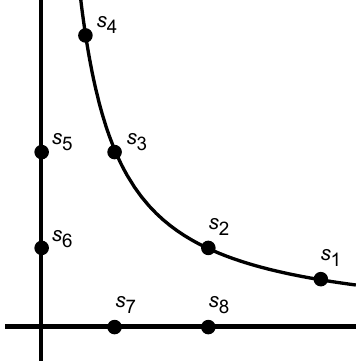}
         \caption{}
     \end{subfigure}
     \hspace{2cm}
     \begin{subfigure}[b]{0.3\textwidth}
         \centering
         \includegraphics[width=\textwidth]{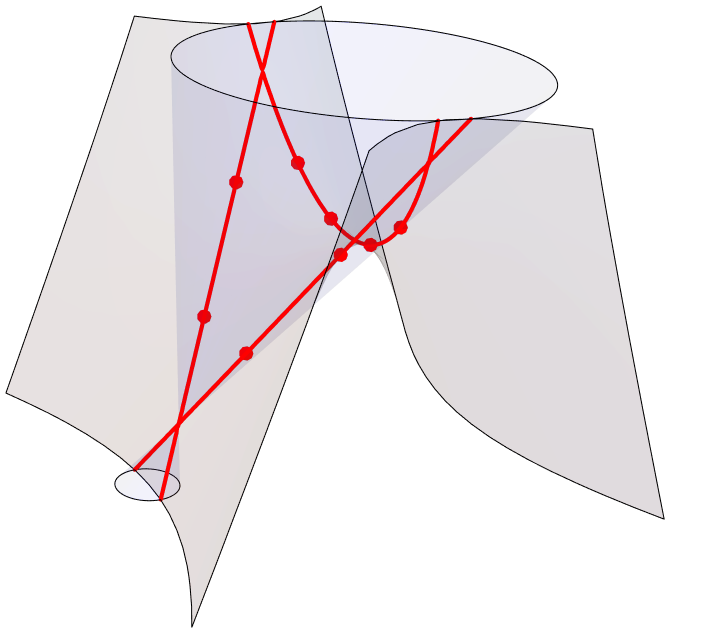}
         \caption{}
     \end{subfigure}
\end{center}
\caption{(a) Base set of the surface type $A_2^{(1)}$: four points on a (1,1)-curve and two pairs of points on two generators through two different points of the curve. (b) Pencil of quadrics through a conic and two coplanar lines meeting the conic at two different points.}
\label{fig qPV}
\end{figure}

\paragraph{2D QRT map.} We start with the QRT map $f=i_1\circ i_2$ for the pencil of biquadratic curves based on eight points
\begin{equation}
s_i=(a_i,a_i^{-1}), \quad i=1,\ldots,4,
\end{equation}
\begin{equation}
s_5=(0,b_5), \quad s_6=(0,b_6), \quad s_7=(a_7,0), \quad s_8=(a_8,0).
\end{equation}
A straightforward computation shows that these points support a pencil of biquadratic curves if and only if the following condition is satisfied:
\begin{equation}\label{qPV QRT condition}
a_1a_2a_3a_4b_5b_6=a_7a_8.
\end{equation}
The pencil through these eight points contains a reducible curve $C_\infty$ with the equation $xy(xy-1)=0$, consisting of the following three irreducible components: 
\begin{equation}\label{qPV C infty}
C_\infty: \quad\{x=0\}\cup\{y=0\}\cup\{xy=1\}.
\end{equation}
This curve is shown on Fig. \ref{fig qPV} (a). 

The vertical and the horizontal switches $i_1$, $i_2$ for the above mentioned pencil are:
\begin{equation}\label{qPV QRT i1}
i_1(x,y)=(x,\t y), \quad {\rm where}\quad \frac{(\t yx-1)(yx-1)}{\t yy}=\frac{\prod_{i=1}^4(x-a_i)}{(x-a_7)(x-a_8)}, 
\end{equation}
\begin{equation}\label{qPV QRT i2}
i_2(x,y)=(\t x,y), \quad {\rm where}\quad \frac{(\t xy-1)(xy-1)}{\t xx}=\frac{\prod_{i=1}^4(y-a_i^{-1})}{(y-b_5)(y-b_6)},
\end{equation}
and the corresponding QRT map is $f= i_1\circ i_2$. Birational involutions $i_1$, $i_2$ on  $\bbP^1\times \bbP^1$ admit four ``short'' singularity confinement patterns \eqref{sing conf QRT i1i2} and four ``long'' singularity confinement patterns of the types \eqref{sing conf QRT i1i2 long y}, \eqref{sing conf QRT i1i2 long x} (two of each type). 

\paragraph{3D Painlev\'e map.} We consider the pencil of quadrics $\{P_\mu\}$, the Segre lift of the pencil of curves $\{C_\mu\}$, and we declare the pencil $Q_\lambda$ to be spanned by $Q_0$ and $P_{\infty}=X_3(X_4-X_3)$:
\begin{equation}\label{qPV pencil}
Q_\lambda=\Big\{X_1X_2-X_3X_4-(\lambda-1) X_3(X_4-X_3)=0\Big\}
\end{equation}
(shift of the parameter $\lambda\to\lambda-1$ is for convenience, to ensure the canonical normalization of ${\rm Sing}(Q)$).
The base set of the pencil $Q_\lambda$ consists of the two lines $\{X_1=X_3=0\}$, $\{X_2=X_3=0\}$, and the conic $\{X_1X_2-X_3X_4=0,\; X_3=X_4\}$, as on Fig. \ref{fig qPV} (b). Intersection of this base set with the base set of the pencil $\{P_\mu\}$ consists of eight points
\begin{align}\label{qPV P1-P8}
& S_i=[a_i:a_i^{-1}:1:1], \quad i=1,\ldots,4, \nonumber\\
& S_5=[0:b_5:0:1], \quad S_6=[0:b_6:0:1], \nonumber\\ 
& S_7=[a_7:0:0:1], \quad S_8=[a_8:0:0:1].
\end{align}
The characteristic polynomial of the pencil $\{Q_\lambda\}$ is $\Delta(\lambda)=\det(M_\lambda)=\lambda^2$, so that ${\rm Sing}(Q)=\{0,\infty\}$.  The 3D QRT involutions $i_1,i_2$ along generators of the pencil $\{Q_\lambda\}$ till the second intersection with $\{P_\mu\}$, and the 3D QRT map $f=i_1\circ i_2$, as described in Definition \ref{def 3D QRT}, are birational maps of $\mathbb P^3$. 

A M\"obius automorphism of $\bbP^1$ fixing ${\rm Sing}(Q)=\{0,\infty\}$ can be taken as 
$\sigma(\lambda)=q\lambda$ with $q\in\bbC\setminus\{0,1\}$.

\begin{theorem}\label{Th1 for qP(A_2)}
The linear projective map on $\bbP^3$ given by
\begin{equation}\label{qPV L}
L:\quad [X_1:X_2:X_3:X_4]\mapsto [X_1:X_2:X_3:q^{-1}(X_4-X_3)+X_3]
\end{equation}
preserves the pencil $\{Q_\lambda\}$ and sends each $Q_\lambda$ to $Q_{q\lambda}$. Moreover, it is a Painlev\'e deformation map: the birational map $\t f=\t i_1\circ \t i_2$ on $\mathbb P^3$ with $\t i_1=L\circ i_1$, $\t i_2=L\circ i_2$ is a 3D Painlev\'e map sending $Q_\lambda$ to $Q_{q^2\lambda}$ with the following singularity confinement patterns:
\begin{itemize}
\item[-] \eqref{sing conf dP f short} \;for $i=1,\ldots,4$, 
\item[-] \eqref{sing conf dP f long y} \;for $(i,j)=(5,6), (6,5)$, 
\item[-]  \eqref{sing conf dP f long x} \;for $(i,j)=(7,8), (8,7)$.
\end{itemize}
Here \begin{align}
& \Phi_i=\{X_3-a_iX_2=0\}, \quad \Psi_i=\{X_3-a_i^{-1}X_1=0\}, \quad i=1,\ldots,4,\\
& \Psi_i=\{X_1X_2-X_3^2-b_iX_1(X_4-X_3)=0\}, \quad i=5,6 \\
& \Phi_i=\{X_1X_2-X_3^2-a_iX_2(X_4-X_3)=0\}, \quad i=7,8.
\end{align}
\end{theorem}
{\bf Proof} is, like for Theorem \ref{Th1 for dP(A_2)}, by a direct computation. In particular, to prove the statement about singularity confinement, one checks that conditions of Proposition \ref{prop Painleve} are satisfied:
\begin{itemize}
\item $L$ fixes the points $S_i$, $i=1,\ldots,4$, and acts on the other four base points as follows:
\begin{align}
& L(S_5)=[0:qb_5:0:1], \quad L(S_6)=[0:qb_6:0:1], \\
& L(S_7)=[qa_7:0:0:1], \quad L(S_8)=[qa_8:0:0:1].
\end{align}
\item $L\circ i_1$ maps $L(S_5)$ to $S_6$ and $L(S_6)$ to $S_5$.
\item $L\circ i_2$ maps $L(S_7)$ to $S_8$ and $L(S_8)$ to $S_7$.
\end{itemize}

We mention also that $L$ fixes the planes $\Psi_i$ and $\Phi_i$ for $i=1,\ldots,4$, maps the quadrics  $\Psi_i$ to analogous quadrics with $b_i\mapsto qb_i$ for $i=5,6$, and maps the quadrics $\Phi_i$ to analogous quadrics with $a_i\mapsto qa_i$ for $i=7,8$. $\blacksquare$

\paragraph{Relation to the $q$-Painlev\'e equation of the surface type $A_2^{(1)}$.} To establish a relation between the map $\t f$ and a $q$-Painlev\'e equation, we start by computing the normalizing transformation of $Q_\lambda$ to the canonical form $Q_0$: 
\begin{equation}\label{qPV pencil norm}
 \begin{bmatrix} X_1 \\ X_2 \\ X_3 \\ X_4\end{bmatrix}=
 \begin{bmatrix} Y_1 \\ Y_2 \\ Y_3 \\ \lambda^{-1}(Y_4+(\lambda-1) Y_3)\end{bmatrix}
 =A_\lambda\begin{bmatrix} Y_1\\Y_2\\Y_3\\Y_4\end{bmatrix}, \quad
A_\lambda=\begin{pmatrix} 1 & 0 & 0 &  0\\ 0 & 1 & 0 & 0 \\  0 & 0 & 1 & 0 \\ 0 & 0 & \lambda^{-1}(\lambda-1) & \lambda^{-1}\end{pmatrix}.                               
 \end{equation}
 This immediately gives the following parametrization of $Q_\lambda$: 
\begin{equation}\label{qPV pencil x to X}
\begin{bmatrix} X_1 \\ X_2 \\ X_3 \\ X_4 \end{bmatrix}
=\begin{bmatrix} x \\ y \\ xy \\ \dfrac{1}{\lambda}+ \dfrac{\lambda-1}{\lambda}xy  \end{bmatrix}=:\phi_\lambda(x,y).
\end{equation}
The pencil-adapted coordinates $(x,y,\lambda)$ on $\bbP^3$ are:
\begin{equation}\label{qPV pencil X to x}
x = \frac{X_3}{X_2}= \frac{1}{\lambda}\Big(\frac{X_1}{X_4}+(\lambda-1)\frac{X_3^2}{X_2X_4}\Big), \quad y = \frac{X_3}{X_1}= \frac{1}{\lambda}\Big(\frac{X_2}{X_4}+(\lambda-1)\frac{X_3^2}{X_1X_4}\Big), \quad \lambda =\frac{X_1X_2-X_3^2}{X_3(X_4-X_3)}.
\end{equation}

In the pencil-adapted coordinates $(x,y,\lambda)$, for each fixed $\lambda$, the intersection curves $Q_\lambda\cap P_\mu$ form the pencil $\phi_\lambda^* P_\mu$ which can be characterized as the pencil of biquadratic curves in $\bbP^1\times\bbP^1$ through the eight points
\begin{align}\label{qPV p1-p8 lambda}
& s_i(\lambda)=(a_i,a_i^{-1}), \quad i=1,\ldots,4, \nonumber\\
& s_5(\lambda)=(0,b_5\lambda^{-1}), \quad s_6(\lambda)=(0,b_6\lambda^{-1}), \nonumber\\
& s_7(\lambda)=(a_7\lambda^{-1},0), \quad s_8(\lambda)=(a_8\lambda^{-1},0),
\end{align}
which correspond to $S_1,\ldots,S_8$ given in \eqref{qPV P1-P8} under the map $\phi_\lambda^{-1}$. The curve $C_\infty(\lambda)$ has the same equation $\{xy(1-xy)=0\}$ as the curve $C_\infty$ and is given by \eqref{qPV C infty}.
Pencil $\phi_\lambda^* P_\mu$ can be obtained from $C_\mu$ by the modification of parameters $b_i\mapsto b_i\lambda^{-1}$, $i=5,6$, and $a_i\mapsto a_i\lambda^{-1}$, $i=7,8$. Therefore, formulas for the involutions $i_1$, $i_2$ restricted to $Q_\lambda$  coincide with the original formulas \eqref{qPV QRT i1}, \eqref{qPV QRT i2}, with the modified parameters:
\begin{equation}\label{qPV 3D i1 on fiber}
i_1|_{Q_\lambda}:\; (x,y)\mapsto (x,\t y), \; {\rm where}\;\;\frac{(\t yx-1)(yx-1)}{\t yy}=\frac{\prod_{i=1}^4(x-a_i)}{(x-a_7\lambda^{-1})(x-a_8\lambda^{-1})}, 
\end{equation}
\begin{equation}\label{qPV 3D i2 on fiber}
i_2|_{Q_\lambda}: \; (x,y)\mapsto (\t x,y), \; {\rm where} \;\;  \frac{(\t xy-1)(xy-1)}{\t xx}=
\frac{\prod_{i=1}^4(y-a_i^{-1})}{(y-b_5\lambda^{-1})(y-b_6\lambda^{-1})}.
\end{equation}
Now we immediately arrive at the following result:
\begin{theorem}
If one parametrizes $Q_\lambda$ by $(x,y)\in\bbP^1\times \bbP^1$ according to \eqref{qPV pencil x to X}, then in coordinates $(x,y,\lambda)$ on $\bbP^3$ the map $\t f:(x_n,y_n,\lambda_{2n})\mapsto(x_{n+1},y_{n+1},\lambda_{2n+2})$ is equivalent to the $q$-Painlev\'e equation of the surface type $A_2^{(1)}$, a system of two non-autonomous difference equations:
\begin{eqnarray}
 \frac{(x_{n+1}y_n-1)(x_ny_n-1)}{x_{n+1}x_n} & = & 
\frac{\prod_{i=1}^4(y_n-a_i^{-1})}{\big(y_n-b_5\lambda_{2n}^{-1}\big)\big(y-b_6\lambda_{2n}^{-1}\big)},
 \label{qPV x} \\
\frac{(y_{n+1}x_{n+1}-1)(y_nx_{n+1}-1)}{y_{n+1}y_n} & = & 
\frac{\prod_{i=1}^4(x_{n+1}-a_i)}{\big(x_{n+1}-a_7\lambda_{2n+1}^{-1}\big)\big(x_{n+1}-a_8\lambda_{2n+1}^{-1}\big)},
\label{qPV y}
\end{eqnarray}
where $\lambda_n=q^n\lambda_0$.
\end{theorem}


\section{From a pencil of type (viii)  to the {\em d}-Painlev\'e equation of the type $A_1^{(1)}$ }
\label{sect case viii}

\begin{figure}[h!]
\begin{center}
\begin{subfigure}[b]{0.3\textwidth}
         \centering
         \includegraphics[width=0.8\textwidth]{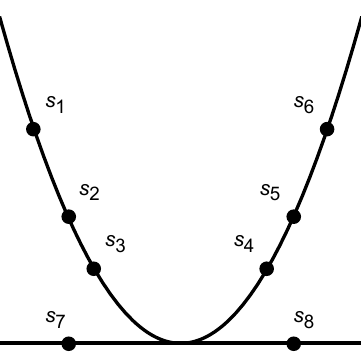}
         \caption{}
     \end{subfigure}
     \hspace{2cm}
     \begin{subfigure}[b]{0.3\textwidth}
         \centering
         \includegraphics[width=\textwidth]{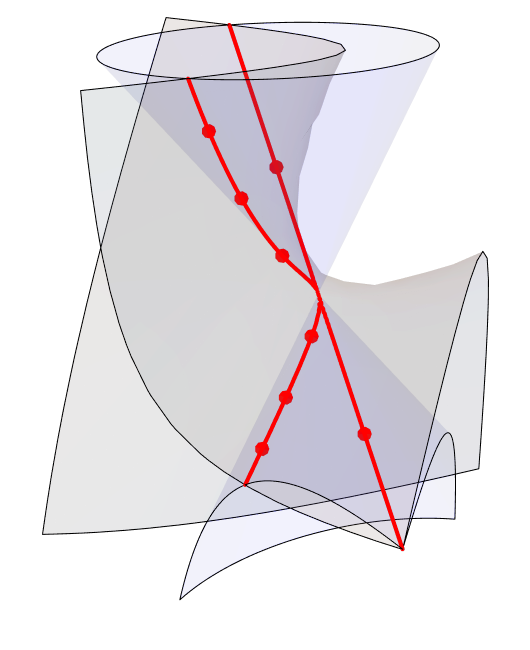}
         \caption{}
     \end{subfigure}
\end{center}
\caption{(a) Base set of the surface type $A_1^{(1)}$: six points on a (2,1)-curve and two points on a (0,1)-curve tangent to it. (b) Pencil of quadrics through a twisted cubic and one of its tangent lines.}
\label{fig case vi}
\end{figure}

\paragraph{2D QRT map.} 
We start with a QRT map corresponding to the pencil of biquadratic curves through the following eight points:
\begin{equation}\label{case vi QRT base points}
s_i=(a_i,a_i^2),\; i=1,\ldots,6, \quad s_7=(a_7,\infty),\quad s_8=(a_8,\infty)
\end{equation}
They lie on a biquadratic curve on $\bbP^1\times\bbP^1$ with equation $\{x^2-y=0\}$. It has two irreducible components:
\begin{equation}\label{case vi planar base curve}
C_\infty: \quad \{y=x^2\}\cup\{y=\infty\},
\end{equation}
a (2,1)-curve and a horizontal generator, see Fig. \ref{fig case vi} (a). One easily checks that the points {case vi QRT base points} support a pencil of biquadraric curves $C_\mu$ (including the curve $C_\infty$) if and only if the following condition is satisfied:
\begin{equation}\label{case vi pencil condition}
a_1+a_2+a_3+a_4+a_5+a_6=a_7+a_8.
\end{equation}

We define $i_1$ and $i_2$ as the vertical and the horizontal switches with respect to this pencil, and we define the QRT map $f=i_1\circ i_2$. The formulas for the involutions $i_1$ and $i_2$ can be written compactly in terms of the following equations:
\begin{equation}\label{case vi QRT i1}
i_1(x,y)=(x,\t{y}), \quad (\t{y}-x^2)(y-x^2)=\frac{\prod_{i=1}^{6}(x-a_i)}{(x-a_7)(x-a_8)},
\end{equation}
\begin{equation}\label{case vi QRT i2}
i_2(x,y)=(\t{x},y),\quad \frac{(\t{x}-\eta)(x-\eta)}{(\t{x}+\eta)(x+\eta)}=\frac{\prod_{i=1}^{6}(\eta-a_i)}{\prod_{i=1}^{6}(\eta+a_i)},\quad y=\eta^2.
\end{equation}
The last equation  \eqref{case vi QRT i2} has to be understood as follows: upon clearing denominators, this becomes the vanishing condition of an odd polynomial of $\eta$ of degree 7. Upon division by $\eta$, this becomes an even polynomial of $\eta$ of degree 6, which is a polynomial of $y=\eta^2$ of degree 3. Thus, it defines $\t x$ as a rational function of $x$ and $y$, of bidegree (1,3). Birational involutions $i_1$, $i_2$ on  $\bbP^1\times \bbP^1$ admit six ``short'' singularity confinement patterns \eqref{sing conf QRT i1i2} and two ``long'' singularity confinement patterns of the type \eqref{sing conf QRT i1i2 long x}.

\paragraph{3D Painlev\'e map.} We consider the pencil of quadrics $\{P_\mu\}$, the Segre lift of the pencil of curves $\{C_\mu\}$, and we declare the pencil $Q_\lambda$ to be spanned by $Q_0$ and $P_{\infty}=X_1^2-X_2X_4$:
\begin{equation}\label{case vi pencil}
Q_\lambda=\big\{X_1X_2-X_3X_4-\lambda(X_1^2-X_2X_4)=0\big\}.
\end{equation}
The base curve of this pencil is:
\begin{equation}\label{case vi base curve}
\big\{[x:x^2:x^3:1]: x\in\bbP^1\big\}\cup\{X_1=X_4=0\} 
\end{equation} 
(the union of a twisted cubic and a tangent line to it). The intersection of this base set with the base set of the pencil $\{P_\mu\}$ consists of eight points
\begin{equation}\label{case vi P1-P8}
S_i=[a_i:a_i^2:a_i^3:1], \; i=1,\ldots,6, \quad S_7=[0:1:a_7:0],\quad S_8=[0:1:a_8:0].
\end{equation}
The characteristic polynomial of the pencil $\{Q_\lambda\}$ is: $\Delta(\lambda)=\det(M_\lambda)=1$, so that ${\rm Sing}(Q)=\{\infty\}$. The 3D QRT involutions $i_1,i_2$ along generators of the pencil $\{Q_\lambda\}$ till the second intersection with $\{P_\mu\}$, and the 3D QRT map $f=i_1\circ i_2$, as described in Definition \ref{def 3D QRT}, are birational maps of $\mathbb P^3$. The explicit formulas for them are omitted.

A M\"obius automorphism of $\bbP^1$ fixing ${\rm Sing}(Q)=\{\infty\}$ can be taken as 
$\sigma(\lambda)=\lambda+\delta$ with $\delta\in\bbC\setminus\{0\}$.

\begin{theorem}\label{Th 1 dE7}
The birational map of degree 2 on $\bbP^3$ given by 
\begin{equation}\label{dE7 L}
L:\begin{bmatrix} X_1\\ X_2 \\ X_3\\ X_4\end{bmatrix}\;\mapsto\; \begin{bmatrix} \hat X_1 \\ \hat X_2\\ \hat X_3\\ \hat X_4\end{bmatrix}=
\begin{bmatrix} X_1X_4 \\  X_2X_4 \\ X_3X_4+\delta(X_2X_4-X_1^2) \\ X_4^2\end{bmatrix},
 \end{equation}
preserves the pencil $\{Q_\lambda\}$ and sends each $Q_\lambda$ to $Q_{\lambda+\delta}$. Moreover, it is a Painlev\'e deformation map: the birational map $\t f=\t i_1\circ \t i_2$ on $\mathbb P^3$ with $\t i_1=L\circ i_1$, $\t i_2=L\circ i_2$ is a 3D Painlev\'e map. It sends $Q_\lambda$ to $Q_{\lambda+2\delta}$  and has, away from $Q_\infty$, the following singularity confinement patterns: 
\begin{itemize}
\item[-] \;\eqref{sing conf dP f short}\; for $i=1,\ldots,6$, 
\item[-]  \;\eqref{sing conf dP f long x}\; for $(i,j)=(7,8), (8,7)$.
\end{itemize}
Here
\begin{align}
& \Phi_i=\{X_1-a_iX_4=0\}, \quad i=1,\ldots,6,\\ 
& \Psi_i=\{X_1X_3-X_2^2-a_i^2(X_1^2-X_2X_4)+a_i(X_1X_2-X_3X_4)=0\}, \quad i=1,\ldots,6,\\
& \Phi_i=\{X_1^3-X_3X_4^2-a_iX_4(X_1^2-X_2X_4)=0\}, \quad i=7,8.
\end{align}
\end{theorem}
{\bf Proof.} Unlike all previous cases, where the Painlev\'e deformation map $L$ was a projective linear map in $\bbP^3$, here it is a birational map of degree 2. Finding this map and verification of the statements of the theorem  only becomes feasible by a fiberwise consideration. Therefore we start with the normalizing transformation of $Q_\lambda$ to the canonical form $Q_0$: 
\begin{equation}\label{case vi pencil norm}
 \begin{bmatrix} X_1 \\ X_2 \\ X_3 \\ X_4\end{bmatrix}=
 \begin{bmatrix} Y_1 \\ Y_2+\lambda Y_1 \\ Y_3+\lambda Y_2+\lambda^2 Y_3 \\ Y_4 \end{bmatrix}
 =A_\lambda\begin{bmatrix} Y_1\\Y_2\\Y_3\\Y_4\end{bmatrix}, \quad
A_\lambda=\begin{pmatrix} 1 & 0 & 0 &  0\\ \lambda & 1 & 0 & 0 \\  \lambda^2 & \lambda & 1 & 0 \\ 0 & 0 & 0 & 1 \end{pmatrix}.                               
 \end{equation}
 This immediately gives the following parametrization of $Q_\lambda$: 
\begin{equation}\label{case vi pencil x to X}
 \left[\begin{array}{c} X_1 \\ X_2\\ X_3\\ X_4\end{array}\right]=
 \left[\begin{array}{c} x \\
                               y+ \lambda x\\ 
                                  xy+\lambda y +\lambda^2 x \\ 
                                  1\end{array}\right]=:\phi_\lambda(x,y).
 \end{equation}
The pencil-adapted coordinates $(x,y,\lambda)$ on $\bbP^3$ are:
\begin{equation}\label{case vi pencil X to x}
x=\frac{X_1}{X_4}=\frac{X_3-\lambda X_2}{X_2-\lambda X_1},\quad y=\frac{X_2-\lambda X_1}{X_4}=\frac{X_3-\lambda X_2}{X_1}, \quad \lambda=\frac{X_1X_2-X_3X_4}{X_1^2-X_2X_4}.
\end{equation}
The curve $C_\infty(\lambda)$ defined by eq. \eqref{base curve lambda} is given by 
\begin{equation}\label{case vi Cinfty lambda}
C_\infty(\lambda): \quad \{y=x^2-\lambda x\}\cup\{y=\infty\}.
\end{equation}
It supports the points
\begin{align}\label{case vi p1-p8 lambda}
& s_i(\lambda)=(a_i,a_i^2-\lambda a_i), \; i=1,\ldots,6, \quad
 s_7(\lambda)=(a_7-\lambda,\infty), \quad s_8(\lambda)=(a_8-\lambda,\infty),
\end{align}
which correspond to $S_1,\ldots,S_8$ given in \eqref{case vi P1-P8} under the map $\phi_\lambda^{-1}$. 

The fiberwise construction \eqref{L on fiber} requires to find $\psi_\lambda:\bbP^1\times\bbP^1\to\bbP^1\times\bbP^1$ which maps the curve $C_\infty(\lambda)$ to $C_\infty(\lambda+\delta)$. In the present example, we should have $\psi_\lambda(x,y)=(\hat{x},\hat{y})$ such that
$$
\hat{y}=\infty\;\Leftrightarrow\;y=\infty, \quad \hat{y}=\hat{x}^2-(\lambda+\delta)\hat{x}\;\Leftrightarrow\;
y=x^2-\lambda x.
$$
To ensure that $\psi_\lambda$ sends $s_i(\lambda)$ to $s_i(\lambda+\delta)$, we should have $\hat x=x$. This leads to
\begin{equation}\label{dE7 psi}
\psi_\lambda:\quad    \hat{x}=x, \quad  \hat{y}=y-\delta x.
\end{equation}
Thus, in the pencil-adapted coordinates $(x,y,\lambda)$ the action of $L$ is described by
\begin{equation}\label{dE7 L  in xy}
L: (x,y,\lambda)\mapsto (x,y-\delta x,\lambda+\delta).
\end{equation}
Now a direct computation with \eqref{case vi pencil x to X} results in \eqref{dE7 L}. We remark that the critical set of $L$ is $\mathcal C(L)=\{X_4=0\}$, while its indeterminacy set is $\mathcal I(L)=\{X_1=X_4=0\}$. By construction, the map $L$ fixes the 
twisted cubic $\{[t:t^2:t^3:1]:\,t\in\bbP^1\}$ pointwise. In particular, it fixes the points $S_i$, $i=1,\ldots,6$.

We now turn to the proof of the statements about the singularity confinement for the map $\t f$. As usual, we refer to Proposition \ref{prop Painleve}. The patterns \eqref{sing conf dP f short} involving $S_i$, $i=1,\ldots,6$ follow from Proposition \ref{prop Painleve} , since these points are fixed by the deformation map $L$. 

Turning to the patterns \eqref{sing conf dP f long x}, we encounter the following problem: they involve expressions $L(S_7)$, $L(S_8)$, which are actually not well-defined, since $S_7,S_8\in\mathcal I(L)$. We will nevertheless, by abuse of notation, assume that
\begin{equation}\label{dE7 L(s_i)}
L(S_7)=[0:1:a_7+\delta:0], \quad 
L(S_8)=[0:1:a_8+\delta:0].
\end{equation}
Let us comment on this. The foliation of $\mathbb P^3$ by the quadrics $Q_\lambda$ gives us effectively a blow-up of the set $\mathcal I (L)$, which is described in the pencil-adapted coordinates as $\{y=\infty\}$, so that the point $[0:1:a:0]\in\mathcal I(L)$ has coordinates $(x,y)=(a-\lambda,\infty)$ on $Q_\lambda$. As long as $\lambda\neq \infty$, any point $(a-\lambda,\infty)\in Q_\lambda$ is mapped to $(a-\lambda,\infty)\in Q_{\lambda+\delta}$ with homogeneous coordinates $[0:1:a+\delta:0]$. For $a=a_7, a_8$, this justifies formulas \eqref{dE7 L(s_i)} away from $Q_\infty$.

To prove \eqref{sing conf dP f long x}, we make the fiberwise computation. On the quadric $Q_\lambda$, $i_2$ maps 
$(x,\infty)$ to $(a_7+a_8-2\lambda-x,\infty)$. Therefore, we have
\begin{align*}
& \{x=a_7-\lambda\}\;{\rm on}\; Q_\lambda\\
& \quad \xrightarrow[]{i_1} (a_7-\lambda,\infty)\;{\rm on}\; Q_\lambda
 \xrightarrow[]{L} (a_7-\lambda,\infty)\;{\rm on}\; Q_{\lambda+\delta}\\
& \quad \xrightarrow[]{i_2}(a_8-(\lambda+2\delta),\infty)\;{\rm on}\; Q_{\lambda+\delta}
\xrightarrow[]{L}(a_8-(\lambda+2\delta),\infty)\;{\rm on}\; Q_{\lambda+2\delta}\\
& \quad \xrightarrow[]{i_1}\{x=a_8-(\lambda+2\delta)\}\;{\rm on}\; Q_{\lambda+2\delta}
\xrightarrow[]{L}\{x=a_8-(\lambda+2\delta)\}\;{\rm on}\; Q_{\lambda+3\delta}.
\end{align*}
This can be written (away from $Q_\infty$) as 
\begin{align}
\Phi_7\xrightarrow[]{\t i_1} L(S_7)\xrightarrow[]{\t i_2}S_8\xrightarrow[]{\t i_1}L(\Phi_8).
\end{align}
In terms of $\t f=\t i_1\circ\t i_2$, this gives \eqref{sing conf dP f long x}, as claimed.  

Equations of the surfaces $\Psi_i$ and $\Phi_i$ are also most easily computed fiberwise, by lifting their traces on the fibers $Q_\lambda$ to homogeneous coordinates with the help of formulas \eqref{case vi pencil X to x}. $\blacksquare$

\paragraph{Relation to the $d$-Painlev\'e equation of the surface type $A_1^{(1)}$.} 

In the pencil-adapted coordinates $(x,y,\lambda)$, for each fixed $\lambda$, the intersection curves $Q_\lambda\cap P_\mu$ form the pencil $\phi_\lambda^* P_\mu$ which can be characterised as the pencil of biquadratic curves in $\bbP^1\times\bbP^1$ through the eight points \eqref{case vi p1-p8 lambda}.
The formulas for the involutions $i_1$, $i_2$ restricted to $Q_\lambda$:
\begin{equation}\label{case vi 3D QRT i1 on fiber} 
i_1|_{Q_\lambda}(x,y)=(x,\t{y}), \quad \big(\t{y}-x(x-\lambda)\big)\big(y-x(x-\lambda)\big)=\frac{\prod_{i=1}^{6}(x-a_i)}
{(x-a_7+\lambda)(x-a_8+\lambda)},
\end{equation}
\begin{equation}\label{case vi 3D QRT i2 on fiber}
i_2|_{Q_\lambda}(x,y)=(\t{x},y),\quad \frac{(\t{x}-\eta)(x-\eta)}{(\t{x}+\eta-\lambda)(x+\eta-\lambda)}=\frac{\prod_{i=1}^{6}(\eta-a_i)}{\prod_{i=1}^{6}(\eta+a_i-\lambda)},\quad y=\eta(\eta-\lambda).
\end{equation}

\begin{theorem}\label{Th dE7}
If one parametrizes $Q_\lambda$ by $(x,y)\in\bbP^1\times \bbP^1$ according to \eqref{case vi pencil x to X}, then in coordinates $(x,y,\lambda)$ on $\bbP^3$ the map $\t f:(x_n,y_n,\lambda_{2n})\mapsto(x_{n+1},y_{n+1},\lambda_{2n+2})$ is equivalent to the $d$-Painlev\'e equation of the surface type $A_1^{(1)}$, the system of two non-autonomous difference equations:
\begin{equation} \label{case vi x}
\frac{(x_{n+1}-\eta_n)(x_n-\eta_n)}{(x_{n+1}+\eta_n-\lambda_{2n})(x_n+\eta_n-\lambda_{2n})}=\frac{\prod_{i=1}^{6}(\eta_n-a_i)}{\prod_{i=1}^{6}(\eta_n+a_i-\lambda_{2n})},\quad y_n=\eta_n(\eta_n-\lambda_{2n}),
\end{equation}
\begin{eqnarray}\label{case vi y}
\lefteqn{\big(y_{n+1}-x_{n+1}(x_{n+1}-\lambda_{2n+2})\big)\big(y_n-x_{n+1}(x_{n+1}-\lambda_{2n})\big)}\nonumber\\
&&
\qquad \qquad=\frac{\prod_{i=1}^{6}(x_{n+1}-a_i)}
{(x_{n+1}-a_7+\lambda_{2n+1})(x_{n+1}-a_8+\lambda_{2n+1})},
\end{eqnarray}
where $\lambda_n=\delta n+\lambda_0$.
\end{theorem}
{\bf Proof.} Equation \eqref{case vi x} is obtained from \eqref{case vi 3D QRT i2 on fiber} by $\lambda\leftarrow \lambda_{2n}$, $x\leftarrow x_n$, $\t x\leftarrow x_{n+1}$, and  $\eta\leftarrow \eta_n$. Similarly, equation \eqref{case vi y} is obtained from \eqref{case vi 3D QRT i1 on fiber} by $\lambda\leftarrow \lambda_{2n+1}$, $x\leftarrow x_{n+1}$, $y\leftarrow y_n-\delta x_{n+1}$, and $\t y\leftarrow y_{n+1}+\delta x_{n+1}$. $\blacksquare$
\medskip

\noindent
{\bf Remark.} As pointed out in the introduction, the discrete Painlev\'e equations of this and the next sections are different from discrete Painlev\'e equations related to the surface type $A_1^{(1)}$ usually encountered in the literature (cf. \cite{KNY}). While the former seems to be new, the latter was recently introduced in \cite{N, NY}. These equations come from a different realization of the surfaces of type $A_1^{(1)}$ than the standard one. Namely, in our realization the roots are $[D_1]=2H_1+H_2-\sum_{i=1}^6 E_i$ and $[D_2]=H_2-E_7-E_8$, i.e., $D_1$ corresponds to a curve of bidegree (2,1) through six blow-up points and $D_2$ corresponds to a horizontal line (a curve of bidegree (0,1)) through two blow-up points. In the realization in \cite{KNY}, the roots are $[D_1]=H_1+H_2-\sum_{i=1}^4 E_i$ and $[D_2]=H_1+H_2-\sum_{i=5}^8 E_i$, i.e., $D_1$ and $D_2$ correspond to two curves of bidegree (1,1) through four blow-up points each. Interestingly, Sakai used in \cite{Sakai} the same realization as ours (up to a canonical birational isomorphism between $\mathbb P^2$ and $\mathbb P^1\times\mathbb P^1$), but he did not give the corresponding discrete Painlev\'e equations explicitly.


\section{From a pencil of type (vii) to the {\em q}-Painlev\'e equation of the surface type $A_1^{(1)}$ }
\label{sect case vii}

\begin{figure}[!ht]
\begin{center}
\begin{subfigure}[b]{0.3\textwidth}
         \centering
         \includegraphics[width=0.8\textwidth]{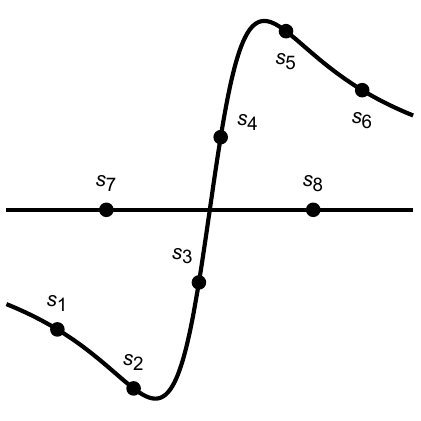}
         \caption{}
         \label{fig:case vii planar}
     \end{subfigure}
     \hspace{2cm}
     \begin{subfigure}[b]{0.33\textwidth}
         \centering
         \includegraphics[width=\textwidth]{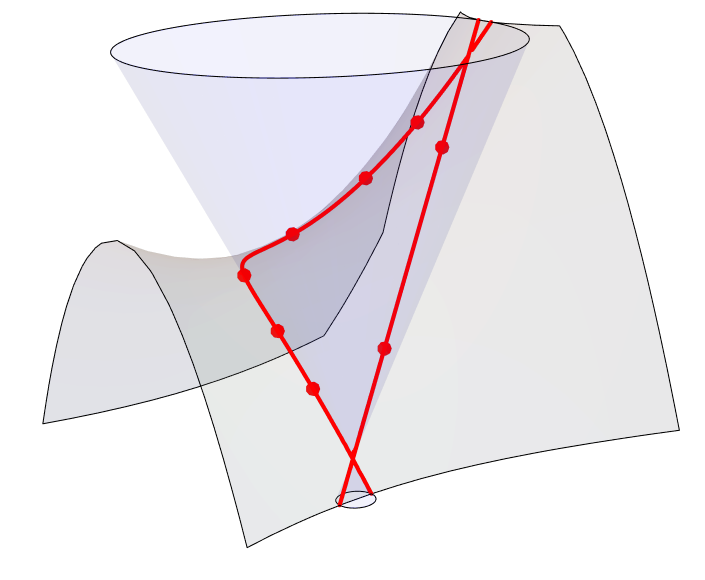}
         \caption{}
     \end{subfigure}
         \label{fig: case vii space}
\end{center}
\caption{(a) Base set of the surface type $A_1^{(1)}$: six points on a (2,1)-curve and two points on a secant (0,1)-curve. (b) Pencil of quadrics through a twisted cubic and one of its secant lines.}
\label{fig case vii}
\end{figure}

\paragraph{2D QRT map.} 
As the last (and the most complicated) example we consider the following configuration of eight points in $\mathbb P^1$:
\begin{equation}\label{case vii QRT base points}
s_i=\big(a_i,a_i+a_i^{-1}\big),\; i=1,\ldots,6, \quad s_7=(a_7,\infty),\quad s_8=(a_8,\infty).
\end{equation}
These eight points lie on a biquadratic curve with the equation $C_\infty: \{x^2+1-xy=0\}$, which has two irreducible components: 
\begin{equation}\label{case vii planar base curve}
C_\infty: \quad \{y=x+x^{-1}\}\cup\{y=\infty\},
\end{equation}
a (2,1)-curve and a (0,1)-curve. See Fig. \ref{fig case vii} (a).
One shows that these points support a pencil of biquadraric curves $C_\mu$ (including the curve $C_\infty$) if and only if the following condition is satisfied:
\begin{equation}\label{case vii pencil condition}
a_1a_2a_3a_4a_5a_6=a_7a_8.
\end{equation}
The vertical and the horizontal switches with respect to this pencil are given by:
\begin{equation}\label{case vii QRT i1}
i_1(x,y)=(x,\t{y}), \quad (x\t{y}-x^2-1)(xy-x^2-1)=\frac{\prod_{i=1}^{6}(x-a_i)}{(x-a_7)(x-a_8)},
\end{equation}
\begin{equation}\label{case vii QRT i2}
i_2(x,y)=(\t{x},y),\quad \eta^2\frac{(\t{x}-\eta)(x-\eta)}{(\t{x}-\eta^{-1})(x-\eta^{-1})}=\eta^{-2}\,\frac{\prod_{i=1}^{6}(\eta-a_i)}{\prod_{i=1}^{6}(\eta^{-1}-a_i)},\quad y=\eta+\eta^{-1}.
\end{equation}
The last equation  \eqref{case vii QRT i2} has to be understood as follows. Clearing denominators, we write it as
$$
\eta^2(\t{x}-\eta)(x-\eta)\prod_{i=1}^{6}(\eta^{-1}-a_i)-\eta^{-2}(\t{x}-\eta^{-1})(x-\eta^{-1})\prod_{i=1}^{6}(\eta-a_i)=0.
$$
The left-hand side is a Laurent polynomial of $\eta$ with terms from $\eta^{-4}$ to $\eta^4$, vanising upon $\eta\mapsto\eta^{-1}$. Upon division by $\eta-\eta^{-1}$, this becomes a Laurent polynomial of $\eta$ with terms from $\eta^{-3}$ to $\eta^3$, symmetric under $\eta\mapsto\eta^{-1}$. Thus, it is a polynomial of $y=\eta+\eta^{-1}$ of degree 3, and defines $\t x$ as a rational function of $x$ and $y$, of bidegree (1,3). 

The birational involutions $i_1$, $i_2$ on  $\bbP^1\times \bbP^1$ admit six ``short'' singularity confinement patterns \eqref{sing conf QRT i1i2} and two ``long'' singularity confinement patterns of the type \eqref{sing conf QRT i1i2 long x}. 

\paragraph{3D Painlev\'e map} We consider the pencil of quadrics $\{P_\mu\}$, the Segre lift of the pencil of curves $\{C_\mu\}$, and we declare the pencil $Q_\lambda$ to be spanned by $Q_0$ and $P_{\infty}$:
\begin{equation}\label{case vii pencil}
Q_\lambda=\big\{X_1X_2-X_3X_4+(\lambda-1) (X_1^2+X_4^2-X_3X_4)=0\big\}
\end{equation}
(with the parameter of the pencil shifted for later convenience).
The base curve of this pencil is:
\begin{equation}\label{case vii base curve}
\big\{[x^2:x^2+1:x^3+x:x]: x\in\bbP^1\big\}\cup\{X_1=X_4=0\},
\end{equation} 
and consists of a twisted cubic and its secant line. The intersection of this base curve with the base curve of the pencil $\{P_\mu\}$ consists of eight points
\begin{equation}\label{case vii P1-P8}
S_i=\big[a_i:a_i+a_i^{-1}:a_i^2+1:1\big], \; i=1,\ldots,6, \quad S_7=[0:1:a_7:0],\quad S_8=[0:1:a_8:0],
\end{equation}
which are images of $s_i$ under the Segre embedding.

The characteristic polynomial of the pencil $\{Q_\lambda\}$ computes to $\Delta(\lambda)=\det(M_\lambda)=\lambda^2$, so that ${\rm Sing}(Q)=\{0,\infty\}$. The 3D QRT involutions $i_1,i_2$ along generators of the pencil $\{Q_\lambda\}$ till the second intersection with $\{P_\mu\}$, and the 3D QRT map $f=i_1\circ i_2$, as described in Definition \ref{def 3D QRT}, are birational maps of $\mathbb P^3$. The explicit formulas for them are omitted.

A M\"obius automorphism of $\mathbb P^1$ fixing the points of ${\rm Sing}(Q)$ can be taken as $
\sigma(\lambda)=q\lambda$ with $q\in\mathbb C\setminus\{0,1\}$.

\begin{theorem}\label{Th 1 qE7}
The birational map $L$ of degree 2 on $\bbP^3$ given by
\begin{equation}\label{qE7 L}
L:\begin{bmatrix} X_1\\ X_2 \\ X_3\\ X_4\end{bmatrix}\;\mapsto\; \begin{bmatrix} \hat X_1 \\ \hat X_2\\ \hat X_3\\ \hat X_4\end{bmatrix}=
\begin{bmatrix} X_1^2 \\  X_1X_2+(1-q)(X_1^2+X_4^2-X_1X_2) \\ X_1X_3 \\ X_1X_4\end{bmatrix},
 \end{equation}
preserves the pencil $\{Q_\lambda\}$ and sends each $Q_\lambda$ to $Q_{q\lambda}$. Moreover, it is a Painlev\'e deformation map: the birational map $\t f=\t i_1\circ \t i_2$ on $\mathbb P^3$ with $\t i_1=L\circ i_1$, $\t i_2=L\circ i_2$ is a 3D Painlev\'e map. It sends $Q_\lambda$ to $Q_{q^2\lambda}$  and has, away from $Q_0$ and $Q_{\infty}$, the following singularity confinement patterns: 
\begin{itemize}
\item[-] \;\eqref{sing conf dP f short}\; for $i=1,\ldots,6$, 
\item[-]  \;\eqref{sing conf dP f long x}\; for $(i,j)=(7,8), (8,7)$.
\end{itemize}
Here
\begin{align}
& \Phi_i=\{X_1-a_iX_4=0\}, \quad i=1,\ldots,6,\\ 
& \Psi_i=\{X_1^2-X_3 X_4+X_4^2-a_i(X_1 X_3-X_2X_3+X_2X_4) 
 + a_i^2(X_1^2-X_1X_2+X_4^2) =0\}, \nonumber \\ &\hspace{12cm} i=1,\ldots,6,\\
& \Phi_i=\{X_1(X_1^2-X_3 X_4+X_4^2)-a_i X_4(X_1^2-X_1 X_2+X_4^2)=0\}, \quad i=7,8.
\end{align}
\end{theorem}
{\bf Proof.} We work again fiberwise, and towards this goal, we start with computing the normalizing transformation of $Q_\lambda$ to the canonical form $Q_0$: 
\begin{equation}\label{qE7 pencil norm}
\begin{bmatrix} X_1\\X_2\\X_3\\X_4\end{bmatrix}= A_\lambda\begin{bmatrix} Y_1 \\ Y_2\\ Y_3\\ Y_4\end{bmatrix}, \quad
A_\lambda=\begin{pmatrix} 1 & 0 & 0 &  0\\ 
                                            1-\lambda & 1 & 0 & 0 \\ 
                                            0 & 0 & \lambda^{-1} & \lambda^{-1}(\lambda-1) \\ 
                                            0 & 0 & 0 & 1\end{pmatrix}.                               
 \end{equation}
This gives the following parametrization of $Q_\lambda$: 
\begin{equation}\label{qE7 pencil x to X}
 \begin{bmatrix} X_1 \\ X_2\\ X_3\\ X_4\end{bmatrix}=A_\lambda\begin{bmatrix} x \\ y\\  xy\\   1\end{bmatrix}
 =\begin{bmatrix} x  \\ y +(1-\lambda)x \\  \lambda^{-1}( xy+\lambda-1)\\ 1\end{bmatrix}=:\phi_\lambda(x,y).
 \end{equation}
The pencil-adapted coordinates $(x,y,\lambda)$ on $\bbP^3$ are:
\begin{equation}\label{qE7 X to x}
   x=\frac{X_1}{X_4}=\frac{\lambda X_3-(\lambda-1) X_4}{X_2+(\lambda-1) X_1}, \qquad
   y=\frac{X_2+(\lambda-1) X_1}{ X_4}= \frac{\lambda X_3-(\lambda-1) X_4}{X_1},
\end{equation}   
\begin{equation}\label{qE7 X to lambda}   
  \lambda=\frac{X_1^2+X_4^2-X_1X_2}{X_1^2 +X_4^2-X_3 X_4}.
\end{equation}
The curve $C_\infty(\lambda)$ is given by 
\begin{equation}\label{case vii Cinfty lambda}
C_\infty(\lambda): \quad \{y=\lambda x+x^{-1}\}\cup\{y=\infty\}.
\end{equation}
It supports the points
\begin{align}\label{case vii p1-p8 lambda}
& s_i(\lambda)=\big(a_i,\lambda a_i+a_i^{-1}\big), \; i=1,\ldots,6, \quad
 s_7(\lambda)=(\lambda a_7,\infty), \quad s_8(\lambda)=(\lambda a_8,\infty),
\end{align}
which correspond to $S_1,\ldots,S_8$ given in \eqref{case vii P1-P8} under the map $\phi_\lambda^{-1}$. 

The fiberwise construction \eqref{L on fiber} requires to find $\psi_\lambda:\bbP^1\times\bbP^1\to\bbP^1\times\bbP^1$ which maps the curve $C_\infty(\lambda)$ to $C_\infty(q\lambda)$. In the present example, we should have $\psi_\lambda(x,y)=(\hat{x},\hat{y})$ such that
$$
\hat{y}=\infty\;\Leftrightarrow\;y=\infty, \quad \frac{\hat{y}-\hat{x}^{-1}}{\hat{x}}=q\lambda\;\Leftrightarrow\;
\frac{y-x^{-1}}{x}=\lambda.
$$
To ensure that $\psi_\lambda$ sends $s_i(\lambda)$ to $s_i(q\lambda)$, we should have $\hat x=x$. This leads to
\begin{equation}\label{qE7 psi}
\psi_\lambda:\quad    \hat{x}=x, \quad  \hat{y}=qy-(q-1)x^{-1}.
\end{equation}
Thus, in the pencil-adapted coordinates $(x,y,\lambda)$ the action of $L$ is described by
\begin{equation}\label{qE7 L  in xy}
L: (x,y,\lambda)\mapsto (x,qy-(q-1)x^{-1},q\lambda).
\end{equation}
Now a direct computation with \eqref{qE7 X to x}, \eqref{qE7 X to lambda} results in \eqref{qE7 L}. We remark that the critical set of $L$ is $\mathcal C(L)=\{X_1=0\}$, while its indeterminacy set is $\mathcal I(L)=\{X_1=X_4=0\}$. By construction, $L$ fixes the twisted cubic pointwise. In particular, it fixes the points $S_i$, $i=1,\ldots,6$.

Like in the case of Section \ref{sect case viii},  the fibration of $\mathbb P^3$ by the quadrics $Q_\lambda$ gives us effectively a blow-up of $\mathcal I(L)$. Straightforward computations show that, away from the two degenerate quadrics $Q_{0}$, $Q_\infty$, the map $L$ acts on the line $\{X_1=X_4=0\}$ according to the formula
$
L: [0:1:a:0]\mapsto [0:1:q^{-1} a:0].
$
Indeed, the point $[0:1:a:0]\in\mathcal I(L)$ has coordinates $(x,y)=(\lambda a,\infty)$ on $Q_\lambda$. As long as $\lambda\neq 0, \infty$, any point $(\lambda a,\infty)\in Q_\lambda$ is mapped to $(\lambda a,\infty)\in Q_{q\lambda}$ with homogeneous coordinates $[0:1:q^{-1} a:0]$. Thus, on all $Q_\lambda$ with $\lambda\notin\{0,\infty\}$, the images $\psi_\lambda(s_i(\lambda))$ are well defined and are given in homogeneous coordinates on $\mathbb P^3$ by $L(S_i)=S_i$, $i=1,\ldots,6$, and
\begin{equation}\label{qE7 L(s_i)}
L(S_7)  =  [0: 1 : q^{-1} a_7:0],\quad
L(S_8) = [0: 1: q^{-1}a_8: 0].
\end{equation}
The end of the proof is parallel to the proof of Theorem \ref{Th 1 dE7}. $\blacksquare$

\paragraph{Relation to the $q$-Painlev\'e equation of the surface type $A_1^{(1)}$.} 

In the pencil-adapted coordinates $(x,y,\lambda)$, for each fixed $\lambda$, the intersection curves $Q_\lambda\cap P_\mu$ form the pencil $\phi_\lambda^* P_\mu$ which can be characterized as the pencil of biquadratic curves in $\bbP^1\times\bbP^1$ through the eight points \eqref{case vii p1-p8 lambda} which correspond to $S_1,\ldots,S_8$ given in \eqref{case vii P1-P8} under the map $\phi_\lambda^{-1}$. For the involutions $i_1$, $i_2$ we have: $i_1|_{Q_\lambda}(x,y)=(x,\t{y})$, resp. $i_2|_{Q_\lambda}(x,y)=(\t{x},y)$, where $\t y$, resp. $\t x$ satisfy the equations
\begin{equation}\label{case vii 3D QRT i1 on fiber}
 (x\t{y}-\lambda x^2-1)(xy-\lambda x^2-1)=\frac{\lambda^2\prod_{i=1}^{6}(x-a_i)}{(x-\lambda a_7)(x-\lambda a_8)},
\end{equation}
\begin{equation}\label{case vii 3D QRT i2 on fiber}
\eta^2 \frac{(\t{x}-\eta)(x-\eta)}{\big(\t{x}-(\lambda\eta)^{-1}\big)\big(x-(\lambda\eta)^{-1}\big)}=(\lambda\eta)^{-2}\,\frac{\prod_{i=1}^{6}(\eta-a_i)}{\prod_{i=1}^{6}\big((\lambda\eta)^{-1}-a_i\big)},\quad y=\lambda\eta+\eta^{-1}.
\end{equation}

\begin{theorem}\label{Th qE7}
If one parametrizes $Q_\lambda$ by $(x,y)\in\bbP^1\times \bbP^1$ according to \eqref{qE7 pencil x to X}, then in coordinates $(x,y,\lambda)$ on $\bbP^3$ the map $\t f:(x_n,y_n,\lambda_{2n})\mapsto(x_{n+1},y_{n+1},\lambda_{2n+2})$, is equivalent to a $q$-Painlev\'e equation of the type $A_1^{(1)}$, the system of two non-autonomous difference equations
\begin{equation}\label{qE7 1}
\eta_n^2 \frac{(x_{n+1}-\eta_n)(x_n-\eta_n)}{\big(x_{n+1}-(\lambda_{2n}\eta_n)^{-1}\big)\big(x_n-(\lambda_{2n}\eta_n)^{-1}\big)}=(\lambda_{2n}\eta_n)^{-2}\,\frac{\prod_{i=1}^{6}(\eta_n-a_i)}{\prod_{i=1}^{6}\big((\lambda_{2n}\eta_n)^{-1}-a_i\big)},\quad y_n=\lambda_{2n}\eta_n+\eta_n^{-1},
\end{equation}
\begin{equation}\label{qE7 2}
 (x_{n+1}y_{n+1}-\lambda_{2n+2} x_{n+1}^2-1)(x_{n+1}y_n-\lambda_{2n} x_{n+1}^2-1)=\frac{\lambda_{2n+1}^2\prod_{i=1}^{6}(x_{n+1}-a_i)}{(x_{n+1}-\lambda_{2n+1} a_7)(x_{n+1}-\lambda_{2n+1} a_8)},
\end{equation}
where $\lambda_n=q^n\lambda_0$.
\end{theorem}
{\bf Proof.} Equation \eqref{qE7 1} is obtained from \eqref{case vii 3D QRT i2 on fiber} by $\lambda\leftarrow \lambda_{2n}$, $x\leftarrow x_n$, $\t x\leftarrow x_{n+1}$, and  $\eta\leftarrow \eta_n$. Equation \eqref{qE7 2} is obtained from \eqref{case vii 3D QRT i1 on fiber} by $\lambda\leftarrow \lambda_{2n+1}$, $x\leftarrow x_{n+1}$, $y\leftarrow qy_n-(q-1) x_{n+1}^{-1}$, and $\t y\leftarrow q^{-1} y_{n+1}+q^{-1}(q-1) x_{n+1}^{-1}$. $\blacksquare$
\medskip

\section{Conclusions}

After having elaborated in detail on the novel geometric scheme including a large portion of discrete Painlev\'e equations, several directions for further investigations can be sketched.
\smallskip

1. For the seven classes of pencils of quadrics considered above, our present construction consists in a Painlev\'e modification of 3D QRT maps, which are defined using involutions along generators of the pencil. However, these are not the only interesting geometric involutions in this context. In \cite{PSWZ}, further classes of Manin involutions were defined for pencils of higher-degree planar curves of genus 1. These novel Manin involutions can be also generalized to dimension 3. For instance, a 3D generalization of the Manin involution $I_{ij}^{(2)}$ for a pencil of quartic curves with two double points can be proposed as follows. Consider two pencils of quadrics $\{Q_\lambda\}$ and $\{P_\mu\}$ sharing one common quadric $P_\infty$, and let $S_i$, $i=1,\ldots,8$ be the base set of the net of quadrics spanned by these two pencils. Fix two indices $i,j\in\{1,\ldots,8\}$. For a generic $X\in \bbP^3$ (not belonging to the base set of either pencil), determine $\lambda,\mu\in\bbP^1$ such that $X\in Q_\lambda\cap P_\mu$; define $I_{i,j}(X)$ to be the fourth intersection point of the curve $Q_\lambda\cap P_\mu$ with the plane through $S_i$, $S_j$, and $X$. Compositions of such involutions $I_{i,j}$ provide us with novel integrable maps on $\mathbb P^3$ preserving the quadrics of both pencils. Their Painlev\'e deformations and their precise place in the corresponding action of the affine Weyl groups are definitely worth a detailed study.  
\smallskip

2. The most urgent problem is to develop a more general scheme capable of the treatment of the six pencils left open in the present paper. One of the main problems to overcome here is the non-rationality (over $\mathbb P^3$) of the corresponding 3D QRT maps. This is achieved in the forthcoming paper \cite{ASWnets}.
\smallskip

3. The main question raised by a referee of the first version of this paper is, whether our construction yields new insights that lie outside Sakai's theory. One obvious novelty is organizing the family of generalized Halphen surfaces into a pencil of quadrics, with fixed (non-moving in $\mathbb P^3$) anti-canonical divisor and blow-up points. The deformation parameter $\delta$ resp.\ $q$ is not fixed by the geometry of the base points but can be chosen freely. What are these novel features good for? The frank answer is: we do not know yet. However, we are confident that the intrinsic beauty of this novel picture will transcend into new approaches to some important questions of the theory of discrete Painlev\'e equations. In particular, it can be anticipated that our scheme provides a natural novel framework for the isomonodromic description of discrete Painlev\'e equations. This is also the subject of our ongoing research.

\paragraph*{Statement on conflict of interests} The authors confirm that there is no conflict of interests.

\paragraph*{Data availability} Data availability is not applicable to this article as no new data were created or analysed in this study.


\end{document}